\documentclass[aps,prd,twocolumn,twoside,superscriptaddress,floatfix, nofootinbib]{revtex4-1}

\usepackage{graphicx}
\usepackage{ifthen,amsmath,amssymb, bm}
\usepackage{graphicx}
\usepackage[dvipsnames]{xcolor}
\usepackage{hyperref}
\usepackage{float}
\usepackage{aas_macros}
\usepackage[export]{adjustbox}
\usepackage{bookmark}

\newcommand{\beq}{\begin{equation}}
\newcommand{\eeq}{\end{equation}}
 
\newcommand{\htwo}{\hspace{2pt}}
\newcommand{\nhat}{\hat{\textbf{n}}}

\newcommand{\mb}{\mathbf}

\begin{document}

\title{Probing cosmic velocities with the pairwise kinematic Sunyaev-Zel'dovich signal in DESI Bright Galaxy Sample DR1 and ACT DR6}

\author{B.~Hadzhiyska}
\email{boryanah@ast.cam.ac.uk}
\affiliation{Institute of Astronomy, Madingley Road, Cambridge, CB3 0HA, UK}
\affiliation{Kavli Institute for Cosmology Cambridge, Madingley Road, Cambridge, CB3 0HA, UK}

\author{Y.~Gong}
\affiliation{Department of Astronomy, Cornell University, Ithaca, NY 14853, USA}

\author{Y.~Hsu}
\affiliation{University Observatory, Faculty of Physics, Ludwig-Maximilians-Universität, Scheinerstr. 1, 81679 Munich, Germany}
\affiliation{Max-Planck-Institut f\"{u}r extraterrestrische Physik\,(MPE), Giessenbachstrasse 1, 85748 Garching bei M\"{u}nchen, Germany}

\author{P.~A.~Gallardo}
\affiliation{Department of Physics and Astronomy, University of Pennsylvania, Philadelphia, PA 19104, USA}

\author{J.~Aguilar}
\affiliation{Lawrence Berkeley National Laboratory, 1 Cyclotron Road, Berkeley, CA 94720, USA}
\author{S.~Ahlen}
\affiliation{Department of Physics, Boston University, 590 Commonwealth Avenue, Boston, MA 02215 USA}
\author{D.~Alonso}
\affiliation{Department of Physics, University of Oxford, Denys Wilkinson Building, Keble Road, Oxford OX1 3RH, UK}
\author{R.~Bean}
\affiliation{Department of Astronomy, Cornell University, Ithaca, NY 14853, USA}
\author{D.~Bianchi}
\affiliation{Dipartimento di Fisica ``Aldo Pontremoli'', Universit\`a degli Studi di Milano, Via Celoria 16, I-20133 Milano, Italy}
\affiliation{INAF-Osservatorio Astronomico di Brera, Via Brera 28, 20122 Milano, Italy}
\author{D.~Brooks}
\affiliation{Department of Physics \& Astronomy, University College London, Gower Street, London, WC1E 6BT, UK}
\author{F.~J.~Castander}
\affiliation{Institut d'Estudis Espacials de Catalunya (IEEC), c/ Esteve Terradas 1, Edifici RDIT, Campus PMT-UPC, 08860 Castelldefels, Spain}
\affiliation{Institute of Space Sciences, ICE-CSIC, Campus UAB, Carrer de Can Magrans s/n, 08913 Bellaterra, Barcelona, Spain}
\author{T.~Claybaugh}
\affiliation{Lawrence Berkeley National Laboratory, 1 Cyclotron Road, Berkeley, CA 94720, USA}
\author{S.~Cole}
\affiliation{Institute for Computational Cosmology, Department of Physics, Durham University, South Road, Durham DH1 3LE, UK}
\author{A.~Cuceu}
\affiliation{Lawrence Berkeley National Laboratory, 1 Cyclotron Road, Berkeley, CA 94720, USA}
\author{A.~de la Macorra}
\affiliation{Instituto de F\'{\i}sica, Universidad Nacional Aut\'{o}noma de M\'{e}xico,  Circuito de la Investigaci\'{o}n Cient\'{\i}fica, Ciudad Univ
ersitaria, Cd. de M\'{e}xico  C.~P.~04510,  M\'{e}xico}
\author{Arjun~Dey}
\affiliation{NSF NOIRLab, 950 N. Cherry Ave., Tucson, AZ 85719, USA}
\author{S.~Ferraro}
\affiliation{Lawrence Berkeley National Laboratory, 1 Cyclotron Road, Berkeley, CA 94720, USA}
\affiliation{University of California, Berkeley, 110 Sproul Hall \#5800 Berkeley, CA 94720, USA}
\author{A.~Font-Ribera}
\affiliation{Institut de F\'{i}sica d’Altes Energies (IFAE), The Barcelona Institute of Science and Technology, Edifici Cn, Campus UAB, 08193, Bellate
rra (Barcelona), Spain}
\author{J.~E.~Forero-Romero}
\affiliation{Departamento de F\'isica, Universidad de los Andes, Cra. 1 No. 18A-10, Edificio Ip, CP 111711, Bogot\'a, Colombia}
\affiliation{Observatorio Astron\'omico, Universidad de los Andes, Cra. 1 No. 18A-10, Edificio H, CP 111711 Bogot\'a, Colombia}
\author{S.~Gontcho A Gontcho}
\affiliation{Lawrence Berkeley National Laboratory, 1 Cyclotron Road, Berkeley, CA 94720, USA}
\affiliation{University of Virginia, Department of Astronomy, Charlottesville, VA 22904, USA}
\author{G.~Gutierrez}
\affiliation{Fermi National Accelerator Laboratory, PO Box 500, Batavia, IL 60510, USA}
\author{J.~Guy}
\affiliation{Lawrence Berkeley National Laboratory, 1 Cyclotron Road, Berkeley, CA 94720, USA}
\author{H.~K.~Herrera-Alcantar}
\affiliation{Institut d'Astrophysique de Paris. 98 bis boulevard Arago. 75014 Paris, France}
\affiliation{IRFU, CEA, Universit\'{e} Paris-Saclay, F-91191 Gif-sur-Yvette, France}
\author{C.~Howlett}
\affiliation{School of Mathematics and Physics, University of Queensland, Brisbane, QLD 4072, Australia}
\author{D.~Huterer}
\affiliation{Department of Physics, University of Michigan, 450 Church Street, Ann Arbor, MI 48109, USA}
\affiliation{University of Michigan, 500 S. State Street, Ann Arbor, MI 48109, USA}
\author{M.~Ishak}
\affiliation{Department of Physics, The University of Texas at Dallas, 800 W. Campbell Rd., Richardson, TX 75080, USA}
\author{R.~Joyce}
\affiliation{NSF NOIRLab, 950 N. Cherry Ave., Tucson, AZ 85719, USA}
\author{T.~Kisner}
\affiliation{Lawrence Berkeley National Laboratory, 1 Cyclotron Road, Berkeley, CA 94720, USA}
\author{A.~Kremin}
\affiliation{Lawrence Berkeley National Laboratory, 1 Cyclotron Road, Berkeley, CA 94720, USA}
\author{M.~Landriau}
\affiliation{Lawrence Berkeley National Laboratory, 1 Cyclotron Road, Berkeley, CA 94720, USA}
\author{L.~Le~Guillou}
\affiliation{Sorbonne Universit\'{e}, CNRS/IN2P3, Laboratoire de Physique Nucl\'{e}aire et de Hautes Energies (LPNHE), FR-75005 Paris, France}
\author{M.~E.~Levi}
\affiliation{Lawrence Berkeley National Laboratory, 1 Cyclotron Road, Berkeley, CA 94720, USA}
\author{M.~Manera}
\affiliation{Departament de F\'{i}sica, Serra H\'{u}nter, Universitat Aut\`{o}noma de Barcelona, 08193 Bellaterra (Barcelona), Spain}
\affiliation{Institut de F\'{i}sica d’Altes Energies (IFAE), The Barcelona Institute of Science and Technology, Edifici Cn, Campus UAB, 08193, Bellate
rra (Barcelona), Spain}
\author{A.~Meisner}
\affiliation{NSF NOIRLab, 950 N. Cherry Ave., Tucson, AZ 85719, USA}
\author{R.~Miquel}
\affiliation{Instituci\'{o} Catalana de Recerca i Estudis Avan\c{c}ats, Passeig de Llu\'{\i}s Companys, 23, 08010 Barcelona, Spain}
\affiliation{Institut de F\'{i}sica d’Altes Energies (IFAE), The Barcelona Institute of Science and Technology, Edifici Cn, Campus UAB, 08193, Bellate
rra (Barcelona), Spain}
\author{K.~Moodley}
\affiliation{Astrophysics Research Centre, School of Mathematics, Statistics and Computer Science, University of KwaZulu-Natal, Durban 4001, South Africa}
\author{T.~Mroczkowski} 
\affiliation{Institute of Space Sciences (CSIC-ICE), Carrer de Can Magrans, s/n, 08193 Cerdanyola del Vallès, Barcelona, Spain}
\author{S.~Nadathur}
\affiliation{Institute of Cosmology and Gravitation, University of Portsmouth, Dennis Sciama Building, Portsmouth, PO1 3FX, UK}
\author{N.~Palanque-Delabrouille}
\affiliation{IRFU, CEA, Universit\'{e} Paris-Saclay, F-91191 Gif-sur-Yvette, France}
\affiliation{Lawrence Berkeley National Laboratory, 1 Cyclotron Road, Berkeley, CA 94720, USA}
\author{W.~J.~Percival}
\affiliation{Department of Physics and Astronomy, University of Waterloo, 200 University Ave W, Waterloo, ON N2L 3G1, Canada}
\affiliation{Perimeter Institute for Theoretical Physics, 31 Caroline St. North, Waterloo, ON N2L 2Y5, Canada}
\affiliation{Waterloo Centre for Astrophysics, University of Waterloo, 200 University Ave W, Waterloo, ON N2L 3G1, Canada}
\author{F.~Prada}
\affiliation{Instituto de Astrof\'{i}sica de Andaluc\'{i}a (CSIC), Glorieta de la Astronom\'{i}a, s/n, E-18008 Granada, Spain}
\author{F.~J.~Qu}
\affiliation{Kavli Institute for Particle Astrophysics and Cosmology, Stanford University, 452 Lomita Mall, Stanford, CA, 94305, USA}
\affiliation{Department of Physics, Stanford University, 382 Via Pueblo Mall, Stanford, CA, 94305, USA}
\author{I.~P\'erez-R\`afols}
\affiliation{Departament de F\'isica, EEBE, Universitat Polit\`ecnica de Catalunya, c/Eduard Maristany 10, 08930 Barcelona, Spain}
\author{B.~Ried~Guachalla}
\affiliation{Department of Physics, Stanford University, Stanford, CA, USA 94305-4085}
\affiliation{Kavli Institute for Particle Astrophysics and Cosmology, 382 Via Pueblo Mall Stanford, CA 94305-4060, USA}
\affiliation{SLAC National Accelerator Laboratory 2575 Sand Hill Road Menlo Park, California 94025, USA}
\author{G.~Rossi}
\affiliation{Department of Physics and Astronomy, Sejong University, 209 Neungdong-ro, Gwangjin-gu, Seoul 05006, Republic of Korea}
\author{E.~Sanchez}
\affiliation{CIEMAT, Avenida Complutense 40, E-28040 Madrid, Spain}
\author{E.~Schaan}
\affiliation{Kavli Institute for Particle Astrophysics and Cosmology,
382 Via Pueblo Mall Stanford, CA 94305-4060, USA}
\affiliation{SLAC National Accelerator Laboratory 2575 Sand Hill Road Menlo Park, California 94025, USA}
\author{D.~Schlegel}
\affiliation{Lawrence Berkeley National Laboratory, 1 Cyclotron Road, Berkeley, CA 94720, USA}
\author{M.~Schubnell}
\affiliation{Department of Physics, University of Michigan, 450 Church Street, Ann Arbor, MI 48109, USA}
\affiliation{University of Michigan, 500 S. State Street, Ann Arbor, MI 48109, USA}
\author{H.~Seo}
\affiliation{Department of Physics \& Astronomy, Ohio University, 139 University Terrace, Athens, OH 45701, USA}
\author{C.~Sif\'on}
\affiliation{Instituto de F\'isica, Pontificia Universidad Cat\'olica de Valpara\'iso, Casilla 4059, Valpara\'iso, Chile}
\author{J.~Silber}
\affiliation{Lawrence Berkeley National Laboratory, 1 Cyclotron Road, Berkeley, CA 94720, USA}
\author{D.~Sprayberry}
\affiliation{NSF NOIRLab, 950 N. Cherry Ave., Tucson, AZ 85719, USA}
\author{G.~Tarl\'{e}}
\affiliation{University of Michigan, 500 S. State Street, Ann Arbor, MI 48109, USA}
\author{E.~M.~Vavagiakis}
\affiliation{Department of Physics, Duke University, Durham, NC 27710, USA}
\author{B.~A.~Weaver}
\affiliation{NSF NOIRLab, 950 N. Cherry Ave., Tucson, AZ 85719, USA}
\author{R.~Zhou}
\affiliation{Lawrence Berkeley National Laboratory, 1 Cyclotron Road, Berkeley, CA 94720, USA}
\author{H.~Zou}
\affiliation{National Astronomical Observatories, Chinese Academy of Sciences, A20 Datun Road, Chaoyang District, Beijing, 100101, P.~R.~China}

\begin{abstract}
We present a measurement of the pairwise kinematic Sunyaev-Zel'dovich (kSZ) signal using the Dark Energy Spectroscopic Instrument (DESI) Bright Galaxy Sample (BGS) Data Release 1 (DR1) galaxy sample overlapping with the Atacama Cosmology Telescope (ACT) CMB temperature map. Our analysis makes use of $1.6$ million galaxies with stellar masses $\log M_\star/M_\odot > 10$, and we explore measurements across a range of aperture sizes ($2.1' < \theta_{\rm ap} < 3.5'$) and stellar mass selections. 
This statistic directly probes the velocity field of the large-scale structure, a unique observable of cosmic dynamics and modified gravity. In particular, at low redshifts, this quantity is especially interesting, as deviations from General Relativity are expected to be largest. Notably, our result represents the highest-significance low-redshift ($z \sim 0.3$) detection of the kSZ pairwise effect yet. In our most optimal configuration ($\theta_{\rm ap} = 3.3'$, $\log M_\star > 11$), we achieve a $5\sigma$ detection. 
Assuming that an estimate of the optical depth and galaxy bias of the sample exists via e.g., external observables, this measurement constrains the fundamental cosmological combination $H_0 f \sigma_8^2$. A key challenge is the degeneracy with the galaxy optical depth. We address this by combining CMB lensing, which allows us to infer the halo mass and galaxy population properties, with hydrodynamical simulation estimates of the mean optical depth, $\bar \tau$. 
We stress that this is a proof-of-concept analysis; with BGS DR2 data we expect to improve the statistical precision by roughly a factor of two, paving the way toward robust tests of modified gravity with kSZ-informed velocity-field measurements at low redshift.
\end{abstract}
\maketitle

\section{Introduction} 
\label{sec:intro}

Understanding the origin of the accelerated expansion of the Universe constitutes one of the most major issues in modern physics. A number of different probes such as the cosmic microwave background (CMB) radiation \citep[e.g.,][]{2020A&A...641A...6P,2013ApJS..208...19H,2013PhRvD..87j3012C,2013ApJ...779...86S}, Baryon acoustic oscillations (BAO) \citep[e.g.,][]{2005ApJ...633..560E,2010MNRAS.401.2148P,2021PhRvD.103h3533A,2024arXiv240403000D} and type Ia supernovae \citep[e.g.,][]{1998AJ....116.1009R,1999ApJ...517..565P} offer a powerful path to constraining the expansion history of the Universe. The standard cosmological model ($\Lambda$CDM) explains the accelerated expansion by assuming General Relativity (GR) as the theory of gravity and adopting a cosmological constant, $\Lambda$, as the dominant form of dark energy  \citep{2003RvMP...75..559P,2015PhR...568....1J}. Given some known issues with this model such as the fine-tuning and coincidence problems \citep{2008JCAP...05..007Q,2011PhRvD..83d3518S,2014EPJC...74.3160V,2017Galax...5...17D}, cosmologists have proposed modifications of gravity, which have also been considered as alternatives to dark energy \citep[see][for a review]{2012PhR...513....1C}. While these modifications can be tuned to match the observed expansion history of the Universe, they also predict differences in the growth of structure and the dynamical properties of galaxies and clusters of galaxies \citep[e.g.,][]{2014JCAP...07..050B,2015Univ....1..123D,2018JCAP...04..005C,2019JCAP...06..020P}. Thus, studying the dynamics of galaxy clusters provides an independent and complementary probe for testing beyond-$\Lambda$CDM extensions.

Galaxy groups and clusters are the largest gravitationally-bound cosmic structures. As such, they contain a massive reservoir of ionized gas, which leaves a distinct imprint on the CMB through the thermal and kinematic Sunyaev-Zel'dovich (tSZ, kSZ) effects. The tSZ effect is associated with the inverse-Compton scattering
of CMB photons off electrons, as they pass through the hot, ionized intra-cluster gas of massive clusters \citep{1970Ap&SS...7....3S,1972CoASP...4..173S,1999PhR...310...97B,2002ARA&A..40..643C}. The kSZ effect is the result of the bulk motion of groups and clusters imparting a Doppler shift to the temperature of the CMB signal \citep{1972CoASP...4..173S,1980MNRAS.190..413S}. While the kSZ effect shifts the CMB blackbody spectrum, it does not distort it, whereas the tSZ effect does impart spectral distortions. Because the thermal electron velocities within the cluster are much larger than its bulk velocity, the tSZ effect dominates, with the amplitude of the kSZ effect being an order of magnitude smaller \citep{1999PhR...310...97B}.

Both effects have been studied at great length. The tSZ effect has been explored both through its contribution to the total CMB temperature power spectrum \citep[e.g.,][]{2014JCAP...04..014D,2015ApJ...799..177G} as well as via individual clusters studies and cross-correlations with other large-scale structure (LSS) surveys \citep[e.g.,][]{2010ApJ...716.1118P,2012NJPh...14b5010B,2013ApJ...768..177S,2021PhRvD.103f3513S,2021PhRvD.104d3503V,2022A&A...660A..27T,2025PhRvD.112d3525L,2025arXiv250607432P}. However, the kSZ signal has proved more challenging to detect due both to its smaller amplitude and its effect on the CMB spectrum, which makes it harder to distinguish from fluctuations in the CMB. Nonetheless, the kSZ effect has great potential to shed light both on astrophysical as well as cosmological conundrums \citep[e.g.,][]{1991ApJ...372...21R,1996MNRAS.279..545H,2005MNRAS.356.1477D,2007ApJ...659L..83B,2008PhRvD..77h3004B}. From an astrophysical point of view, the kSZ signal can be used to probe so-called `missing baryons' \citep[e.g.,][]{2015PhRvL.115s1301H,2016PhRvD..93h2002S}, as it is sensitive to baryons that reside in the otherwise hard-to-study diffuse, warm-hot intergalactic medium \citep[e.g.,][]{2008IAUS..244..136M}. Conversely, peculiar velocities encoded in the kSZ effect contain valuable information about the amplitude and growth rate of density perturbations, which as discussed can help elucidate the nature of dark energy and modified gravity \citep{2013ApJ...765L..32K,2014PhLB..735..402M,2015PhRvD..92f3501M,2016PhRvD..93f4026B,2022GReGr..54...44S}.

On scales much larger than individual groups and cluster (of the order of $\lesssim$ 100 Mpc), the gravitational attraction between galaxy groups and clusters causes them, on average, to fall towards each other. A pairwise correlation statistic that folds in the differences of measured temperatures on the sky at the locations of observed galaxies and clusters can be constructed to extract this signal. This means pairwise cluster momentum estimator, or pairwise kSZ estimator for short, was first detected \citep{2012PhRvL.109d1101H} using the Atacama Cosmology Telescope (ACT) DR4 CMB data \citep{2011ApJS..194...41S} and the LSS survey of the Sloan Digital Sky Survey (SDSS) \citep{1998AJ....116.3040G,2006AJ....131.2332G}. Subsequent measurements covering larger survey areas and greater numbers of sources have also been made \citep{2017JCAP...03..008D,2021PhRvD.104d3502C}. Furthermore, the pairwise kSZ signal has been detected using CMB temperature maps from the Planck collaboration with SDSS \citep{2016A&A...586A.140P} and the South Pole Telescope collaboration (SPT) with a cluster catalog from the Dark Energy Survey (DES) \citep{2011PASP..123..568C,2016MNRAS.461.3172S,2023PhRvD.107d2004S}. Other techniques for measuring the kSZ signal include projected fields \citep{2016PhRvL.117e1301H,2016PhRvD..94l3526F,2021PhRvD.104d3518K,2023JCAP...03..039B}, velocity reconstruction (kSZ stacking) \citep{2014MNRAS.443.2311L,2016PhRvD..93h2002S,2021PhRvD.103f3513S,2022A&A...662A..48T,2023PhRvD.108b3516M,2024arXiv240707152H,2025PhRvD.111b3534H,2025arXiv250319870R}, kSZ tomography \citep{2018arXiv181013423S,2021PhRvD.103h3519S}, individual cluster measurements \citep{2013ApJ...778...52S,2018JCAP...02..032M}, CMB temperature dispersion \citep{2018A&A...617A..48P}, 21cm intensity mapping \citep{2019PhRvD.100b3517L}, signatures from CMB anisotropies \citep{2015ApJ...799..177G}, and Fourier space analysis \citep{2024ApJS..271...30L}. See Ref.~\citep{2018arXiv181013423S} for a comprehensive comparison between the different kSZ estimators.

While this statistic, the pairwise kSZ signal, is robust to unwanted contamination from the tSZ and dust emission, it is sensitive to both the cluster peculiar velocity and the integrated optical depth, complicating the disentanglement of its astrophysical and cosmological yield.
Therefore, in order to interpret the signal, it is helpful to combine the pairwise kSZ measurement with an estimate of the optical depths of the same group or cluster sample \citep{2010ApJ...725.1452S,2016JCAP...08..058B}. The optical depth can be estimated by using additional probes that are sensitive to the baryon content such as the tSZ effect and X-ray observations. However, even then, obtaining the needed quantity, i.e., mean optical depth, still presents challenges and potential biases, as one needs to correctly model other complex gas properties (temperature, metallicity, etc.) \citep{2013MNRAS.432.1600D,2017ApJ...837..124F,2018JCAP...02..032M,2021PhRvD.104d3502C,2023PhRvD.107d2004S,2021PhRvD.104d3503V,2024PhRvD.109b3513G}. Alternatively, one can make use of state-of-the-art hydrodynamical simulations, relying on their accuracy to reproduce the correct feedback properties observed in the Universe. This turns out to also present a number of challenges: the main ones being galaxy sample selection, simulation resolution, and feedback strength and (subgrid) implementation \citep{2021PhRvD.103f3514A,2023MNRAS.526..369H,2024arXiv240707152H,2024MNRAS.534..655B,2025MNRAS.540..143M}.

In this work, we present a measurement of the pairwise velocity momentum signal around the Dark Energy Spectroscopic Instrument (DESI) Bright Galaxy Sample (BGS) using the kSZ effect from the ACT latest data release (DR6). Interestingly, a number of dark energy and modified gravity models show the largest deviations from the standard cosmological model ($\Lambda$CDM) at low redshifts. Therefore, measurements using the lowest-redshift tracer, BGS, give a unique window to testing these models through the peculiar velocity of galaxy groups and clusters. For this reason, we also perform a restricted cosmological analysis presenting constraints on the growth rate of structure and expansion history of the Universe. This paper is part of a forthcoming series of analyses combining DESI DR1 and ACT DR6 data to measure the pairwise kSZ and tSZ effects using different galaxy and cluster tracers \citep[in prep.]{Gong2025,Hsu2025,Moore2025}.

This paper is organized as follows. In Section~\ref{sec:theory}, we introduce the kSZ pairwise estimator and detail our approach to modeling the optical depth of the BGS host halos. Next, we describe the data sets used in this study in Section~\ref{sec:data}. In Section~\ref{sec:results}, we present our results of the measurement and interpretation. Furthermore, we place a simple constraint on the cosmological parameters, which our statistic of interest is sensitive to. Finally, in Section~\ref{sec:conc}, we summarize our main findings and comment on future directions. 

\begin{figure*}
\centering
\includegraphics[width=\textwidth]{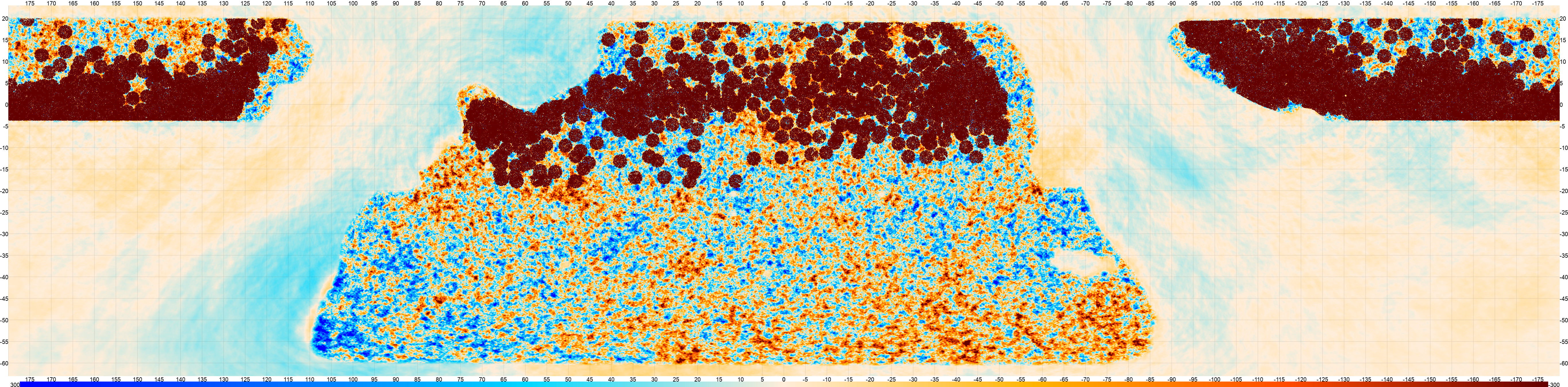}
\caption{Right ascension and declination distribution of BGS Y1 galaxies with $\log_{10}(M_\star / M_\odot) > 10$ in the ACT footprint. 
In total, the sample contains 1,610,381 galaxies. 
Future DESI data releases, such as Y3, are expected to increase the available number of galaxies in this footprint by roughly a factor of four, further improving the statistical precision of kSZ measurements.}
\label{fig:radec}
\end{figure*}

\begin{figure}
\centering
\includegraphics[width=0.5\textwidth]{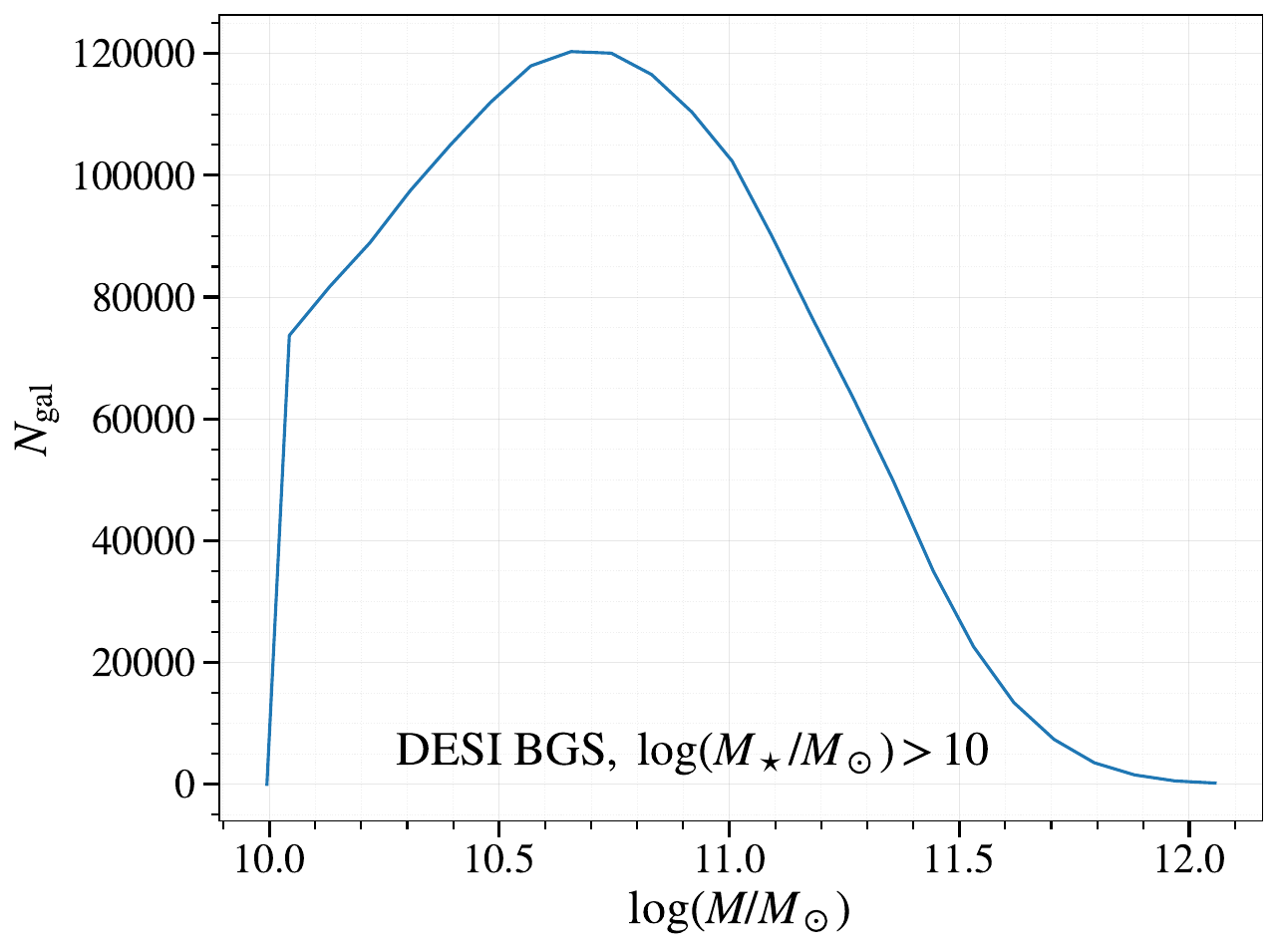}
\includegraphics[width=0.5\textwidth]{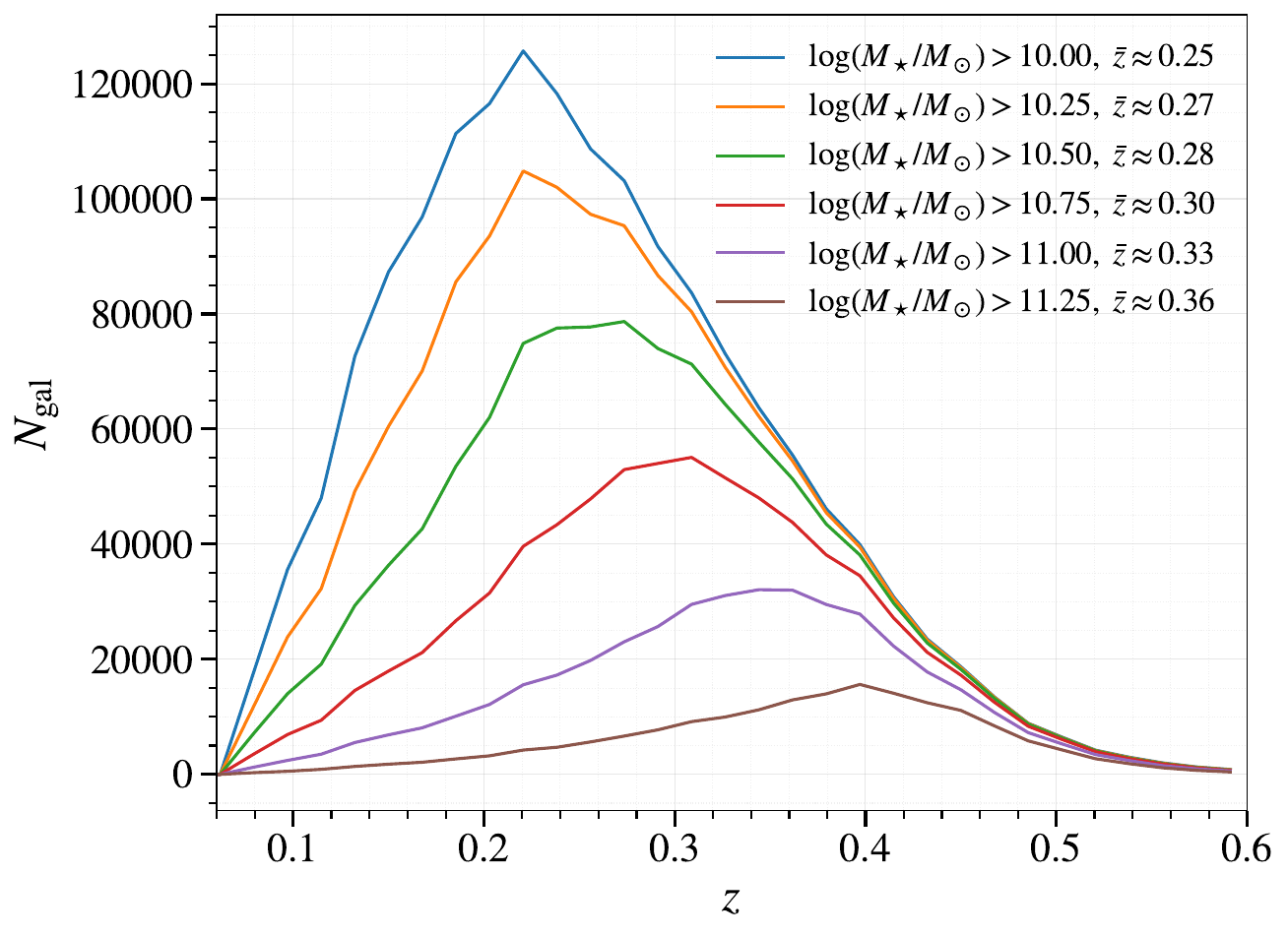}
\caption{{\it Top panel:} Number of BGS galaxies, with a stellar mass threshold of $\log_{10}(M_\star / M_\odot) > 10$, in Y1 as a function of stellar mass in differential bins. The distribution peaks near $\log_{10}(M_\star / M_\odot) \approx 10.75$, reflecting the stellar-mass completeness of the sample at the median survey redshift. {\it Bottom panel:} Number of BGS galaxies per redshift bin for six cumulative stellar-mass thresholds. Higher-mass samples trace progressively higher-redshift populations, with the mean redshift increasing from $\bar{z} \approx 0.25$ at $\log_{10}(M_\star / M_\odot) > 10$ to $\bar{z} \approx 0.36$ at $\log_{10}(M_\star / M_\odot) > 11.25$. These trends highlight the trade-off between number density and redshift reach when selecting stellar-mass subsamples. The number of galaxies in each mass bin is given in Table~\ref{tab:mass}.}
\label{fig:logm}
\end{figure}

\section{Data}
\label{sec:data}

\subsection{Dark Energy Spectroscopic Instrument}

DESI is a highly multiplexed, robotically fibre-positioned spectroscopic array installed on the Mayall 4-meter telescope at Kitt Peak National Observatory \citep{2016arXiv161100036D,DESI2022.KP1.Instr}, capable of obtaining nearly 5000 simultaneous spectra in a three-degree field-of-view \citep{2023AJ....165....9S,Corrector.Miller.2023}. Each of the ten spectrographs covers the entire UV-to-near-IR spectral range \citep{DESI2016b.Instr}. 
Currently, DESI is conducting a five-year survey spanning a 14,000 deg$^2$ sky footprint \citep{FiberSystem.Poppett.2024,Spectro.Pipeline.Guy.2023}, designed to yield a total of approximately 40 million galaxy and quasar spectra and measure cosmological parameters to sub-percent precision over $0 < z < 3.5$ \citep{SurveyOps.Schlafly.2023}.

The target selection pipeline for spectroscopic follow-up is based on imaging data from the public DESI Legacy Imaging Surveys \citep{2017PASP..129f4101Z,2019AJ....157..168D}. Targets are selected via a combination of quality cuts, colour selections and machine learning classification tools, tailored to provide a complete and clean sample for each tracer type. The identification and prioritization of the targets is described in \citep{2023AJ....165...50M}. Of most relevance to this study is the Bright Galaxy Survey \citep[BGS][]{2020RNAAS...4..187R}.

We base our analysis on the \texttt{BGS\_BRIGHT\_full} catalog from DESI DR1 \citep{DESI2024.I.DR1,DESI2024.VII.KP7B,tr6y-kpc6}, which contains all Main Survey targets meeting the BGS-BRIGHT selection criteria.
We further match this catalog to the DESI DR1 value-added catalog (VAC) of physical properties derived with the \textsc{CIGALE} code (v.22.1) \cite{2024A&A...691A.308S,2025arXiv250609143S},
\textsc{CIGALE} models galaxy spectral energy distributions (SEDs) from the optical to infrared by combining stellar population synthesis, 
delayed star formation histories with optional bursts, nebular emission, dust attenuation, 
dust re-emission, 
and AGN emission, fitting simultaneously to photometry in the $g,r,z,W1-W4$ bands.  
Physical properties such as stellar mass, star formation rate, and AGN fraction are estimated via a Bayesian likelihood analysis, with uncertainties given by the standard deviation of the probability distribution functions. The catalog assumes a WMAP7 cosmology and includes $\sim$17 million galaxies across all primary target types: BGS, luminous red galaxies (LRGs), emission-line galaxies (ELGs), and quasars (QSO).  

Our sample effectively adopts the same quality and type cuts on the BGS catalog as the CIGALE outputs: \texttt{COADD\_FIBERSTATUS} = 0, \texttt{ZWARN} = 0 or 4, \texttt{ZCAT\_PRIMARY} = \texttt{True}, and \texttt{SPECTYPE} = GALAXY or QSO. These cuts select objects with reliable spectra and redshifts as well as available photometry from the imaging survey Wide-field Infrared Survey Explorer (WISE). 
Our final analysis therefore uses BGS-BRIGHT galaxies with stellar masses and other physical properties taken directly from the DR1 VAC, \url{https://data.desi.lbl.gov/doc/releases/dr1/vac/cigale/}.

In Fig.~\ref{fig:radec}, we show the angular distribution of the BGS Y1 galaxies in the ACT footprint for $\log_{10}(M_\star / M_\odot) > 10$, corresponding to a total of 1.61 million objects. This is already a large statistical sample, and the forthcoming Y3 data will increase the galaxy numbers in the overlapping footprint by about a factor of four. 

The top of Fig.~\ref{fig:logm} presents the differential stellar-mass distribution of BGS Y1 galaxies. The number counts peak near $\log_{10}(M_\star / M_\odot) \approx 10.75$, a reflection of both the intrinsic stellar-mass function and the completeness limits of the sample at its typical redshifts. The bottom panel illustrates the redshift distributions for six cumulative stellar-mass thresholds. As expected, the higher-mass selections yield progressively higher mean redshifts, from $\bar{z} \approx 0.25$ at $\log_{10}(M_\star / M_\odot) > 10$ to $\bar{z} \approx 0.36$ at $\log_{10}(M_\star / M_\odot) > 11.25$. This clearly demonstrates the trade-off between number density and stellar mass reach that will shape the sensitivity of our kSZ analysis.

\subsection{Atacama Cosmology Telescope}

This paper employs the
harmonic-space Internal Linear Combination (hILC) 
maps \cite{2024PhRvD.109f3530C} from the Data Release 6 (DR6) from the Atacama Cosmology Telescope (ACT), a 6m telescope located in the Atacama Desert of Chile, which measured the CMB from 2007 to 2022. The DR6 data comprise multifrequency observations from 2017 to 2022 across roughly a third of the sky at three frequency bands: \texttt{f090}, \texttt{f150}, and \texttt{f220}. The observational program of DR6 targeted the ``wide'' field. 
For this work, we use only the night-time portion of the data taken in the first five observing seasons.
The ACT maps are produced in the Plate-Carree (CAR) projection scheme. This analysis uses the first version of the ACT DR6 maps, dr6.01. 

 As in Ref.~\cite{2025arXiv250208850L}, we apply a mask on the ACT map that removes all detected clusters, bright point sources, unobserved regions and the Galactic plane.
Since the ACT map features an inpainting of the point sources, which can potentially remove the subtle SZ signal we are looking for, we also mask the inpainted regions. Namely, we cut out 3$'$ discs around subtracted sources, 6$'$ around very bright point sources, and 10$'$ around bright extended sources (for more details, see Ref.~\cite{2024PhRvD.109f3530C}). Additionally, massive clusters dominate the SZ signal (especially the tSZ, which can be a significant contaminant to the pairwise signal) and can bias our kSZ measurements. Thus, to reduce the variance at little cost to the signal, we also mask the ACT DR6 clusters with ${\rm SNR} > 6$ \citep{2025arXiv250721459A}.

Throughout this work we adopt the hILC map. We verify in Appendix~\ref{app:hilc_vs_single} that the kSZ sigal is consistent across hILC and the single-frequency maps, \texttt{f090} and \texttt{f150}. The agreement between \texttt{f090} and \texttt{f150} shows that the pairwise estimator is robust to foregrounds.

\section{Theory}
\label{sec:theory}

\subsection{Basic intuition}
\label{sec:intuition}

The kSZ effect produced by a galaxy group or cluster with position on the sky $\mathbf{\hat n}_i$
results in a change in the observed CMB temperature $T_{\mathrm{CMB}}$ given by
\beq
\frac{\Delta T}{T_{\mathrm{CMB}}} (\nhat_i) = - \tau_{e,i} (\nhat_i) \frac{\nhat_i \cdot \mathbf{v}_i}{c} \, ,
\label{eq:ksz}
\eeq
where we have assumed that we are in the non-relativistic regime and that individual photons scatter only once \citep{1980MNRAS.190..413S}. Here, $\nhat_i \cdot \mathbf{v}_i$ is the projection of the group or cluster velocity $\mathbf{v}_i$ along the line of sight $\nhat_i$, $c$ is the speed of light, and $\tau_{e,i}$ is the Thomson optical depth given by the line-of-sight integral of the free electron number density $n_{e,i}$,
\beq
\tau_{e,i}(\nhat)  = \int  \mathrm{d}l\htwo  n_{e,i}(\nhat, l) \sigma_T \, ,
\eeq
where $\sigma_T$ is the Thomson cross section. The kSZ effect creates a particular pattern in the CMB, consisting of temperature increments and decrements at the locations of clusters moving towards and away from the observer, respectively \citep{2000ApJ...533L..71D}. Thus, it probes the bulk momentum of the ionized gas surrounding a galaxy group or cluster, projected along the line of sight.

Because the kSZ signal has the same spectral shape as the primary CMB and its amplitude is small compared to e.g., the tSZ amplitude, measuring the imprint of individual groups and clusters can be even more challenging \citep[see][for an exception]{2013ApJ...778...52S,2017A&A...598A.115A,2024ApJ...968...74S}, for this reason we typically study this elusive probe by averaging over tens of thousands of groups and clusters. The statistic of interest to this study is the pairwise kSZ signal, which makes use of the fact that on scales smaller than the homogeneity scale, groups and clusters, on average, fall towards each other due to gravity \citep[e.g.,][]{2007ApJ...659L..83B,2008PhRvD..77h3004B}. 

While the kSZ signal from one individual group or cluster is sensitive to its line-of-sight velocity, the amplitude of the pairwise kSZ signal can be related to the mean relative velocity $v_{12}(r)$ of the groups or clusters. The pairwise kSZ amplitude can thus be written as:
\beq
T_{\mathrm{pkSZ}}(r) \equiv \bar{\tau}_e \, \frac{v_{12}(r)}{c} \, T_{\mathrm{CMB}} \, ,
\label{eq:templ}
\eeq
where $\bar{\tau}_e$ is the average optical depth of the groups or clusters in the sample. Negative values ($T_{\mathrm{pkSZ}} < 0$) corresponds to groups and clusters that are on average approaching each other ($v_{12} <0$), and positive ones ($T_{\mathrm{pkSZ}} > 0$) indicates that they are being pulled away from each other.

\subsection{Pairwise kSZ estimator}
\label{sec:est}

In practice, we calculate the pairwise signal, $v_{12}(r)$, using the estimator proposed in \citep{1999ApJ...515L...1F} by summing over all group or cluster pairs with $i<j$ and distances $|\mb{r}_{ij}| \equiv |\mb{r}_i - \mb{r}_j|$:
\begin{eqnarray}
 & \hat{v}_{12}(r) = & \frac{\sum_{i<j,r} (\hat{\mathbf{r}}_i \cdot \mathbf{v}_i -\hat{\mathbf{r}}_j \cdot \mathbf{v}_j ) \, c_{ij}} {\sum_{i<j,r} c_{ij}^2} \, , \hspace{6pt} \\ & c_{ij} = & \hat{\mb{r}}_{ij} \cdot \frac{\hat{\mb{r}}_i + \hat{\mb{r}}_j}{2} \, ,
\label{eq:ferreira}
\end{eqnarray}
where $c_{ij}$ is a geometrical factor that accounts for the projection along the line-of-sight.

Since we cannot directly measure the velocity field, we instead work with the redshift-evolution-corrected temperature decrement measured around each galaxy, $T(\nhat_i)$ \citep{2012PhRvL.109d1101H}:
\beq
\hat{T}_{\mathrm{pkSZ}}(r) = - \frac{\sum_{i<j, r} \left[ T(\nhat_i)-T(\nhat_j) \right ] c_{ij}}{\sum_{i<j,r} c_{ij}^2} \,.
\label{eq:pkszest}
\eeq
We measure the raw temperature decrements, $\hat T_0$, around each galaxy by adopting a compensated aperture photometry (CAP) filter. In detail, we make a square cutout of the CMB temperature map centered at the RA and DEC of each galaxy and define a disk of some fixed aperture radius, $\theta_{\rm ap}$, and a ring around it of equal area, $\theta_{\rm ap} < \theta < \sqrt{2} \theta_{\rm ap}$. We then measure the mean temperature within both disk and ring and take the difference between the two. This procedure is meant to remove contributions from uncorrelated structure, as well as cancel the long wavelength mode contributions from the primary CMB. For tests and validations of the CAP filter in simulations, see Ref.~\citep{2023MNRAS.526..369H}.

A key advantage of this estimator is that residuals from the primary CMB, astrophysical foregrounds such as the tSZ and CIB, and instrumental noise are uncorrelated with the positions of groups and clusters, and therefore average out in the measurement. A drawback, however, is that applying the estimator over a wide redshift range can introduce biases from redshift-dependent foregrounds. For example, since our galaxy selection is based on a stellar-mass threshold, the mean halo mass of the sample evolves with redshift, which in turn changes correlated contributions such as the tSZ signal and the average optical depth. To facilitate comparison across different samples, we adopt a fixed angular aperture photometry scale (with variations considered to assess the signal-to-noise response for different mass thresholds). We note that this choice means that the aperture corresponds to different physical radii at different redshifts, so even for a sample with fixed halo mass the measured temperature decrements would still exhibit a redshift dependence.

To account for these redshift-dependent effects and obtain an unbiased estimate of the pairwise kSZ signal, we calculate the mean measured temperature as a function of redshift and subtract it from the raw temperature decrements, $\hat{T}_0(\nhat_i)$,  as follows \citep{2012PhRvL.109d1101H}:
\beq
\label{eq:T_zevolcorr}
T(\nhat_i) = \hat{T}_0(\nhat_i) - \frac{ \sum_j \hat{T}_0(\nhat_j) \, G(z_i,z_j,\Sigma_z)}{\sum_j G(z_i,z_j,\Sigma_z)} \, .
\eeq
The second term denotes the redshift-averaged temperature at some redshift, $z_i$, and is calculated from the sum of contributions of groups and clusters at redshift $z_j$ weighted by a Gaussian kernel \mbox{$G(z_i,z_j,\Sigma_z) = \exp \left [-(z_i - z_j)^2/ \left(2 \Sigma_z^2 \right)  \right]$}. Similarly to \citep{2021PhRvD.104d3502C}, we choose $\Sigma_z=0.01$. 
We ascertain that our results are insensitive to this choice by also checking $\Sigma_z=0.005, \ 0.05$.

\subsection{Theoretical model}
\label{sec:model}
Galaxy groups and clusters are located at the peaks of the cosmic density field in (nearly) virialized dark matter structures, known as halos. While the kSZ signal is more sensitive to the halo velocity field, we conduct our measurement using the galaxies, which similarly to the halos are biased tracers, but have unique features such as thermal velocities and non-trivial spatial distribution within the halo. Below we summarize the linear theory derivation for the pairwise kSZ signal obtained from a galaxy sample. 

The galaxy overdensity is defined as:
\beq
    \delta_g (\mathbf{x}) \equiv n(\mathbf{x}) / \bar n - 1 \,,
\eeq
where $n(\mathbf{x})$ is the number density of galaxies and $\bar n$ their mean density.
In linear theory, under the assumption of local bias \citep{1993ApJ...413..447F}, we can relate the galaxy overdensity to the matter overdensity, $\delta (\mathbf{x}) \equiv \rho(\mathbf{x}) / \bar \rho - 1$ where $\rho(\mathbf{x})$ is the matter density and $\bar \rho$ its mean, via:
\beq
    \delta_g (\mathbf{x}) = b_g \, \delta (\mathbf{x}) \, ,
\eeq
where $b_g$ is the linear galaxy bias.

On the other hand, the apparent velocity,  $\mathbf{v} (\mathbf{x}) $ of a galaxy group or cluster consists of a Hubble flow component and a peculiar velocity component, i.e., $\mathbf{v} (\mathbf{x}) = a H \mathbf{x} + \mathbf{u} (\mathbf{x})$, where $a$ is the scale factor 
and {\it H} is the Hubble rate at the redshift of interest. In linear perturbation theory, the velocity field is completely described by its divergence $\vartheta (\mathbf{x}) \equiv \nabla \cdot \mathbf{u} (\mathbf{x}) $, and the density and velocity fields are related simply as \citep{2002PhR...367....1B}:
\beq
 \vartheta  (\mathbf{x}) = -  \dot \delta (\mathbf{x})  =  - a H f \delta (\mathbf{x}) \, ,
\label{eq:thetadelta}
\eeq
where the dot denotes a derivative with respect to conformal time and $f$ is the growth rate of density perturbations defined as \mbox{$f \equiv \mathrm{d} \ln D/ \mathrm{d} \ln a $}, with $D$ the linear growth factor.

Under the assumption that the density and velocity fields are Gaussian, their statistical properties are fully specified by their two-point statistics, which we can define in configuration and Fourier space, respectively, as follows (and analogously for the velocity field, swapping $\delta$ for $\vartheta$):
\beq
\xi (r) \equiv \langle \delta (\mathbf{x}) \, \delta (\mathbf{x} + \mathbf{r})  \rangle
\eeq
\beq
\langle \delta (\mathbf{k}) \, \delta (\mathbf{k}')\rangle = (2 \pi)^3 \delta_D (\mathbf{k} + \mathbf{k}') P(k) \,,
\eeq
where $\delta_D$ is the Dirac delta distribution. Using Eq.~\ref{eq:thetadelta}, we can express the density-velocity correlation function as \citep{2016MNRAS.461.3172S}:
\begin{eqnarray}
\xi^{\delta v}(r,a) & \equiv & \langle \delta(\mb{x}) \, \hat{\mb{r}} \cdot \mb{v}(\mb{x}+\mb{r}) \rangle = \nonumber \\ &=& -\frac{aHf}{2\pi^2} \int_0^\infty \mathrm{d}k\, kP(k,a) j_1(kr) \, ,
\label{eq:xi_deltav}
\end{eqnarray}
where $j_1$ is the order-1 spherical Bessel function of the first kind. In configuration space, we can also write  $\xi^{\delta v}(r,a) = - arHf \bar{\xi}(r,a)/3$, where $\bar{\xi}(r)$ is the correlation function averaged within some sphere of radius $r$. Since the mean pairwise velocity of halos, $v_{12}(r)$, measures the relative velocities of biased density peaks (i.e., their pairwise momenta normalized by their mass), we can write it as \citep{1977ApJS...34..425D,1980lssu.book.....P}:
\begin{align}
v_{12}(r) & = \frac{\langle n(\mb{x}_1) \, n(\mb{x}_2) \, \hat{\mb{r}} \cdot (\mb{v}_2-\mb{v}_1) \rangle }{\langle n(\mb{x}_1) \, n(\mb{x}_2) \rangle} \, ,
\label{eq:v12}
\end{align}
where $n(\mb{x})$ is the number density of halos at position $\mb{x}$.
If we assume that the velocities of halos are unbiased, which holds well for low- and intermediate-mass halos (the focus of this study) and even for massive halos with pair separations of $r \gtrsim 50$ Mpc \citep{2015PhRvD..92l3507B}, we can further simplify this as:
\beq
v_{12}(r) = \frac{ \langle [1 + b_g \delta(\mb{x}_1) ] \, [1 + b_g \delta(\mb{x}_2)] \, \hat{\mb{r}} \cdot (\mb{v}_2-\mb{v}_1)   \rangle }  {1+\xi_g (r)} \, ,
\label{eq:v12xi}
\eeq
where $\xi_g(r)$ is the halo auto-correlation function.
In practice, we observe galaxies hosted in halos across a wide range of masses and thus, we need to evaluate the mass-averaged bias, $b_g \equiv b$, defined as:
\beq
b \equiv \frac{ \int_{M_\mathrm{min}}^{M_\mathrm{max}} \mathrm{d} M \, M \, n(M) \, N_g(M) \, b_h(M)}{ \int_{M_\mathrm{min}}^{M_\mathrm{max}} \mathrm{d} M \, M \, n(M)} \, N_g(M) \, ,
\label{eq:mwbias}
\eeq
where $N_g(M)$ is the halo occupation distribution (HOD) of that sample, $n(M)$ the halo mass function, 
and $b_h(M)$ the halo mass-bias relation. In practice, we can evaluate the linear bias, $b$, directly, e.g. from cross-correlations with CMB lensing, as done in this work.
Neglecting three-point correlation terms in Eq.~\ref{eq:v12xi} and assuming that the higher halo bias moments can be approximated as \mbox{$\xi_{g} \approx b^2 \, \xi$}, we obtain:
\beq
v_{12} (r,a) \simeq \frac{2 \, b \, \xi^{\delta v}(r,a)}{1 + b^2 \, \xi(r,a)} \, .
\label{eq:v12_nonlin}
\eeq

Substituting with the configuration-space expression for Eq.~\ref{eq:xi_deltav}, the mean pairwise streaming velocity becomes \citep{2001MNRAS.326..463S,2007ApJ...659L..83B,2015ApJ...808...47M}
\beq
v_{12}(r,a) \simeq -\frac{2}{3} a \, r \, H \, f \, \frac{b \bar{\xi}(r,a)}{1+b^2\xi(r,a)} \, ,
\label{eq:v12_halo}
\eeq
where $\bar{\xi}(r,a)$ is the matter correlation function. We evaluate $\xi$ and $\bar \xi$ using 
linear theory predictions from the \texttt{colossus} \citep{Diemer2018} package, assuming \textit{Planck} 2018 cosmology.

\begin{figure}
    \centering
    \includegraphics[width=0.48\textwidth]{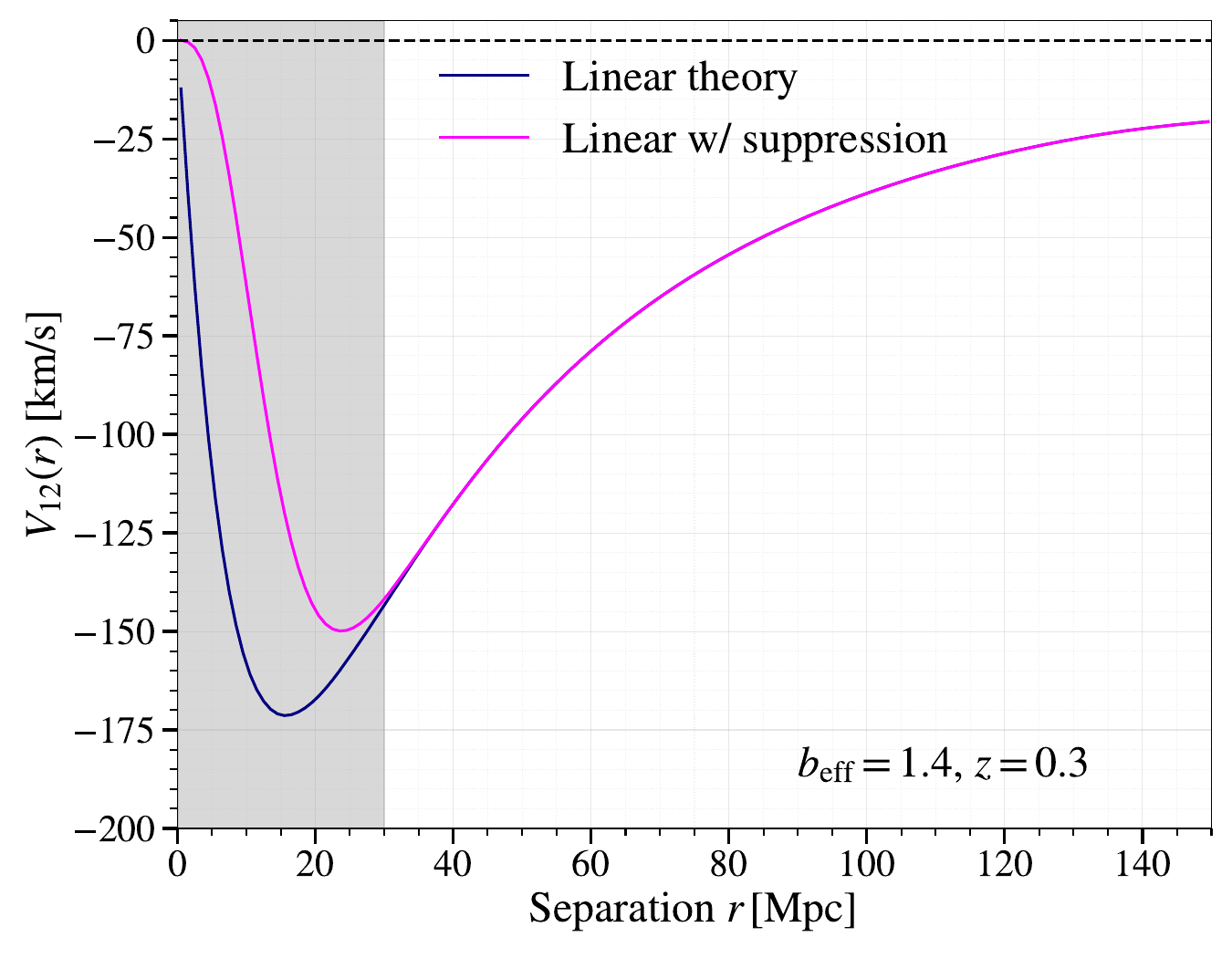}
    \caption{
        Theoretical prediction for the pairwise velocity correlation function, evaluated using 
        the \texttt{colossus} package \citep{Diemer18} with the \textit{Planck} 2018 cosmology. 
        The calculation is performed in linear theory, and thus is only expected to be accurate 
        on intermediate and large scales ($r \gtrsim 30$--$40 \, \mathrm{Mpc}$). 
        On smaller scales, the correlation is additionally suppressed by the thermal motions of galaxies, 
        which smear structure along the line of sight. This effect can be approximated by a damping factor 
        $1 - \exp[-r^2/(2\sigma_r^2)]$, with $\sigma_r \lesssim 10 \, \mathrm{Mpc}$, corresponding to 
        the redshift-space distortion (RSD) smoothing scale. In comparison to the massive clusters studied 
        in Ref.~\citep{2016MNRAS.461.3172S}, where nonlinearities and higher velocity dispersions are more significant, 
        linear theory provides a better description for the less massive galaxy populations studied here. 
        Throughout this work, we therefore focus on scales $r > 30 \, \mathrm{Mpc}$ where the theory is 
        expected to be robust.
    }
    \label{fig:pairwise_theory}
\end{figure}

The pairwise velocity correlation function is shown in Fig.~\ref{fig:pairwise_theory} and evaluated at $z=0.3$, adopting the \textit{Planck} 2018 cosmological parameters. We note two important caveats: (i) linear theory only applies on intermediate and larger scales 
($r \gtrsim 30$--$40 \, \mathrm{Mpc}$), and (ii) thermal velocities of galaxies further damp the 
signal on small scales, an effect that can be modeled by a scale-dependent suppression factor
\begin{equation}
    S(r) = 1 - \exp\left[-\frac{r^2}{2\sigma_r^2}\right],
\end{equation}
with $\sigma_r \lesssim 10 \, \mathrm{Mpc}$, which roughly corresponds to the smearing scale along-the-line of sight due to the thermal motions of galaxies (i.e., Finger-of-God effect) within massive halos (with dispersion velocities of around 600 km/s). 

We show the effects of this suppression in Fig.~\ref{fig:pairwise_theory}. It is evident that even when assuming a large velocity dispersion of $\sim$600 km/s, the changes to the velocity correlation function above 30 Mpc are negligible, which in turn justifies our use of linear theory beyond $r \gtrsim 30 \, \mathrm{Mpc}$ when fitting our measurements. We note that compared to cluster-scale studies \citep[e.g.][]{2016MNRAS.461.3172S}, the lower halo masses in our sample in reality lead to smaller velocity dispersions. In this figure, we assume that the linear bias of the sample is $b \approx 1.4$, which roughly corresponds to our $\log (M_\star/M_\odot) > 11$ sample. When comparing theory and observations, we adopt the best-fit value of $b$ for each sample we consider.


We note that as can be seen from Eq.~\ref{eq:v12_nonlin}, the pairwise kSZ signal constrains a combination of group/cluster astrophysics ($b \bar{\tau}_e$) and cosmology ($\xi^{\delta v}$, $\xi$). Thus, if the cosmological parameters and bias of the sample are fixed, a measurement of the pairwise signal can constrain the mean optical depth of the sample. More relevant to this work is the flipside argument: if external estimates of the  mass-weighted bias $b$  and mean optical depth $\bar{\tau}_e$ were available, one can obtain cosmological constraints. The galaxy bias may be characterised from the auto-correlation function of the galaxy sample used, and $\bar{\tau}_e$ may be extracted from  patchy screening, X-ray or tSZ observations. On large scales ($r \gtrsim 30$ Mpc):
\beq
T_{\mathrm{pkSZ}}(r) \propto b \, \bar{\tau}_e \, \xi^{\delta v}(r) \propto b \, \bar{\tau}_e \, H_0 \, f \, \sigma_8^2 \,,
\label{eq:taubfs82}
\eeq
where we have made the approximation that the term $b^2\xi(r) \ll 1$ can be neglected. We see that the shape of the signal is thoroughly specified by the cosmology through $\xi^{\delta v}(r)$, which is proportional to the combination $H_0 f \sigma_8^2$. This is different from other dynamical probes like redshift space distortions, which are sensitive to the combination $f \sigma_8$ \citep{2009MNRAS.393..297P}. Throughout this study, we set $H_0$ to the fiducial $Planck$ 2018 value. Constraints on $H_0$ require a joint analysis with other cosmological probes  (e.g.\ redshift-space distortions, CMB, weak lensing) that can break these degeneracies, as well as assumptions about the optical depth $\tau_e$. Thus, the pairwise signal can be used to break the degeneracy between growth and primordial amplitude. This could potentially yield powerful measurements of $f$, which can be used as a probe of dark energy and modified gravity \citep{2013ApJ...765L..32K}.

\section{Results}
\label{sec:results}

In the following sections, we present our main findings. We first present our constraints on the linear bias and mean mass of each sample considered in this work. Then we show our measurements of the pairwise kSZ effect and validation of those. Finally, we conclude with implications on the growth of structure constraints.

\subsection{Bias and mass estimates}
\label{sec:bias}

We adopt the procedure introduced in Ref.\ \citep{2025arXiv250714136H} to estimate the mean halo mass and linear bias of DESI BGS Y1 galaxy samples from measurements of the convergence profile $\kappa(\theta)$ using the ACT DR6 CMB lensing map \citep{2024ApJ...962..112Q,2024ApJ...966..138M,2024ApJ...962..113M}.

We obtain the $\hat{\kappa}(\theta)$ profiles in the data by stacking at the positions of all galaxies in our sample. Specifically, we use the \texttt{reproject\_pixell} function to extract cutouts from the DR6 CMB lensing map \citep{2024ApJ...962..112Q}, and then compute the mean signal. The resulting stacked maps are azimuthally averaged in four linearly spaced radial bins between 0 and $6.67'$, yielding the radial $\hat{\kappa}(\theta)$ profiles. 

In brief, we use a vanilla Halo Occupation Distribution (HOD) model applied to the AbacusSummit simulations at fixed redshift $z=0.3$ \citep{2021MNRAS.508.4017M}, appropriate for BGS. The HOD is described by five free parameters, sampled using a Latin Hypercube with 1000 realizations as in the initial work. 

We adopt the \textsc{AbacusHOD} prescription within the \texttt{abacusutils} package \footnote{\url{https://github.com/abacusorg/abacusutils}}. The five free parameters are:
\begin{itemize}
    \item $M_{\mathrm{cut}}$: the characteristic halo mass at which a halo has a 50\% probability of hosting a central galaxy.
    \item $\sigma_{\log M}$: the scatter in $\log M$ describing the smooth transition of the central galaxy occupation function.
    \item $\kappa M_{\rm cut}$: the cutoff mass below which no satellites are hosted.
    \item $M_1$: the typical halo mass required to host one satellite galaxy.
    \item $\alpha$: the power-law slope governing the number of satellites in high-mass halos.
\end{itemize}
The halo mass definition adopted in the \textsc{AbacusSummit} simulations is the `virial mass' one from Ref.~\citep{1998ApJ...495...80B}. 

For each realization, we compute the galaxy--matter cross power spectrum $P_{gm}(k)$, which is then transformed into the CMB convergence profile $\kappa(\theta)$ by performing the Limber integral. We construct an emulator of $\kappa(\theta)$, which was shown in Ref.\ \citep{2025arXiv250714136H} to achieve sub-percent accuracy. We then employ the \texttt{dynesty} nested sampler \citep{2020MNRAS.493.3132S} to constrain the HOD parameters, and from these constraints we derive the mean halo mass and linear bias of each galaxy sample as natural outputs of the AbacusSummit simulation. 

The measured $\kappa(\theta)$ profiles are primarily sensitive to the linear bias and the transition between the one- and two-halo regimes. Consequently, most of the constraining power comes from the mean halo mass and bias. The cross-correlation between galaxies and CMB lensing, used in this work, does not allow us to tightly constrain the full set of HOD parameters by itself. To also achieve tight constraints on the HOD parameters, one needs to consider additional statistics such as the two-point auto-correlation function, which can help break degeneracies between those. For more information, please refer to Ref.~\citep{2025arXiv250714136H}.

We divide the BGS sample into six cumulative stellar-mass thresholds: 
\(\log_{10}(M_\star / M_\odot) > 10, \ 10.25, \ 10.5, \ 10.75, \ 11,\) and \(11.25\). 
For each sample we measure $\kappa(\theta)$ and fit the model described above. The inferred mean masses, linear biases, and $\chi^2_{\rm null}$ values for each stellar-mass threshold are presented in Table~\ref{tab:bgs_bias_mass}. As expected, the linear bias increases with stellar-mass threshold, reflecting the nearly monotonic relation between galaxy stellar mass and host halo mass. This trend is consistent with theoretical expectations and previous observational studies.

\begin{table*}[ht]
\begin{tabular}{lcccccc}
\hline
Param. & ${\log M_\star > 10}$ & ${\log M_\star > 10.25}$ & ${\log M_\star > 10.5}$ & ${\log M_\star > 10.75}$ & ${\log M_\star > 11}$ & ${\log M_\star > 11.25}$ \\
\hline
$\log M_{\rm cut}$ & $11.861^{+0.468}_{-0.543}$ & $11.918^{+0.5}_{-0.599}$ & $11.831^{+0.611}_{-0.607}$ & $11.75^{+0.706}_{-0.519}$ & $11.947^{+0.617}_{-0.46}$ & $12.183^{+0.49}_{-0.418}$ \\ \vspace{0.2cm}
$\log M_1$ & $13.257^{+0.34}_{-0.661}$ & $13.186^{+0.403}_{-0.731}$ & $13.024^{+0.502}_{-0.754}$ & $12.802^{+0.66}_{-0.564}$ & $12.769^{+0.652}_{-0.516}$ & $12.823^{+0.58}_{-0.566}$ \\ \vspace{0.2cm}
$\sigma_{\log M}$ & $0.534^{+0.32}_{-0.356}$ & $0.536^{+0.32}_{-0.359}$ & $0.516^{+0.325}_{-0.343}$ & $0.524^{+0.321}_{-0.347}$ & $0.55^{+0.309}_{-0.369}$ & $0.56^{+0.309}_{-0.367}$ \\ \vspace{0.2cm}
$\alpha$ & $1.21^{+0.504}_{-0.512}$ & $1.21^{+0.485}_{-0.499}$ & $1.322^{+0.395}_{-0.509}$ & $1.358^{+0.351}_{-0.512}$ & $1.315^{+0.374}_{-0.508}$ & $1.346^{+0.354}_{-0.516}$ \\ \vspace{0.2cm}
$\kappa$ & $1.254^{+0.837}_{-0.85}$ & $1.194^{+0.873}_{-0.819}$ & $1.208^{+0.865}_{-0.819}$ & $1.251^{+0.837}_{-0.832}$ & $1.244^{+0.864}_{-0.855}$ & $1.268^{+0.828}_{-0.845}$ \\
\hline
$f_{\rm sat}$ & $0.123^{+0.182}_{-0.053}$ & $0.147^{+0.23}_{-0.069}$ & $0.206^{+0.271}_{-0.105}$ & $0.295^{+0.204}_{-0.169}$ & $0.333^{+0.176}_{-0.18}$ & $0.343^{+0.195}_{-0.166}$ \\ \vspace{0.2cm}
$\log \bar{M}_{\rm h}$ & $13.135^{+0.063}_{-0.103}$ & $13.184^{+0.065}_{-0.105}$ & $13.266^{+0.052}_{-0.088}$ & $13.31^{+0.056}_{-0.071}$ & $13.37^{+0.065}_{-0.086}$ & $13.456^{+0.073}_{-0.117}$ \\ \vspace{0.2cm}
$b_{\rm lin}$ & $1.155^{+0.05}_{-0.035}$ & $1.19^{+0.063}_{-0.042}$ & $1.258^{+0.058}_{-0.051}$ & $1.314^{+0.054}_{-0.05}$ & $1.384^{+0.075}_{-0.07}$ & $1.475^{+0.098}_{-0.108}$ \\
\hline
$\chi^2_{\rm null}$ & 391.121 & 372.607 & 362.447 & 292.895 & 206.465 & 119.811 \\
\hline
\end{tabular}
\caption{Summary of the BGS stellar-mass threshold samples, including values of the 5 HOD parameters, inferred satellite fraction, mean halo mass, linear bias, and $\chi^2_{\rm null}$ with 4 degrees of freedom. Error bars denote $1\sigma$ uncertainties. As estimated in Ref.~\citep{2025arXiv250714136H}, the systematic and model errors expected from this type of analysis are $\sim$7\%.}
\label{tab:bgs_bias_mass}
\end{table*}

\begin{figure}[ht]
    \centering
    \includegraphics[width=0.48\textwidth]{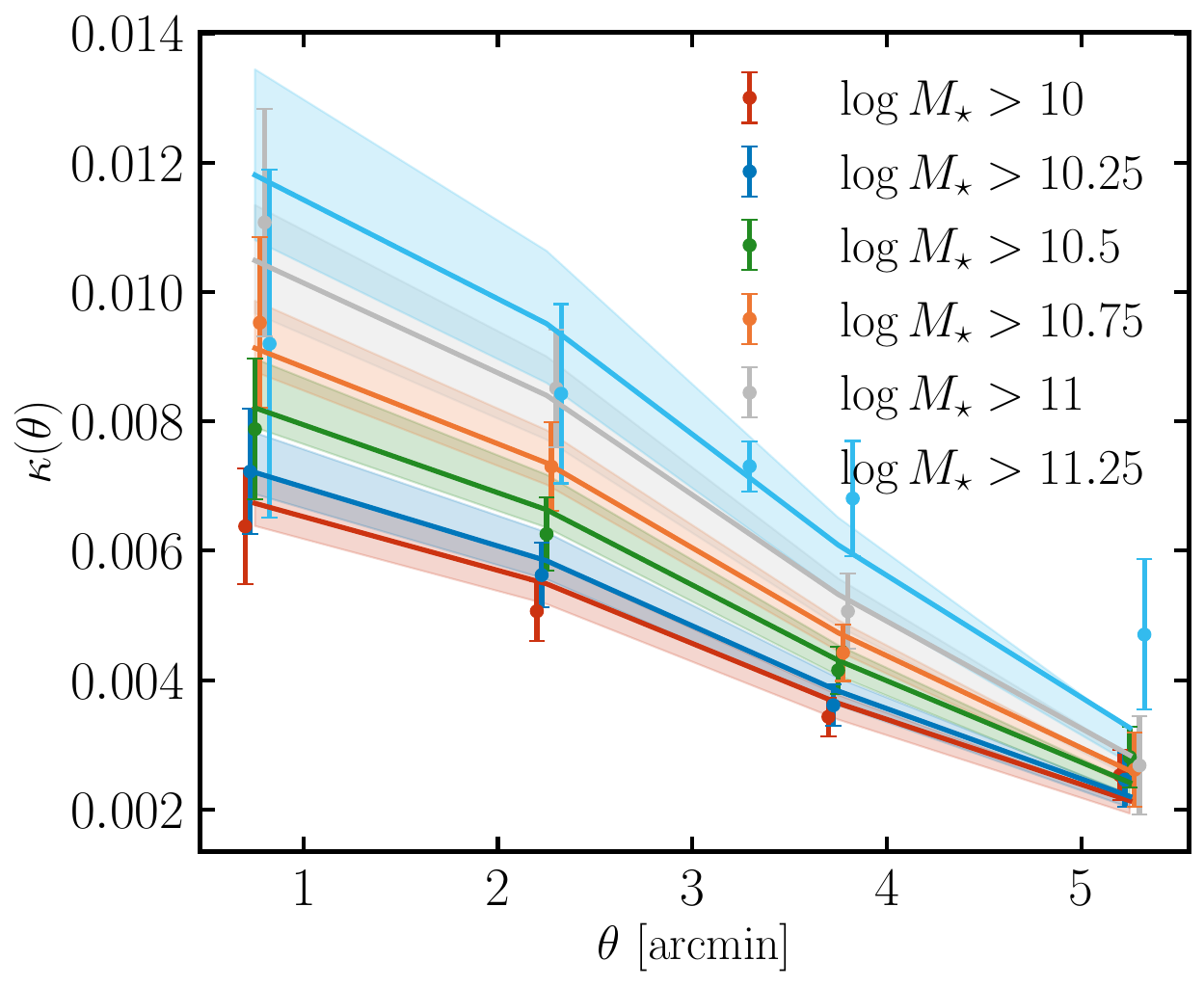}
    \caption{Measured $\kappa(\theta)$ profiles (points) and best-fit theory curves (lines) for the six BGS stellar-mass threshold samples. The amplitude of $\kappa(\theta)$ increases with stellar-mass threshold, consistent with the expectation that more massive galaxies reside in more massive halos with denser matter profiles. The theory curve is obtained using an emulator built on the five standard HOD parameters (see Section~\ref{sec:bias} for more details).}
    \label{fig:kappa_profiles}
\end{figure}

Fig.~\ref{fig:kappa_profiles} shows the measured $\kappa(\theta)$ profiles for the six BGS stellar-mass threshold samples compared with the best-fit theoretical predictions. We see that the amplitude of the $\kappa(\theta)$ profile increases steadily with stellar-mass threshold. This is expected since $\kappa$ directly probes the projected matter density profile, and the amplitude naturally grows with halo mass, which correlates with galaxy stellar mass. The agreement between the measurements and the model across scales demonstrates that our emulator-based inference accurately captures the galaxy-matter connection for these samples. The only sample for which the fit is evidently poor ($\chi^2/{\rm dof} > 1$) is the highest mass sample, with $\log (M_\star/M_\odot) > 11.25$, which suffers from two main issues: number of galaxies in it is small and therefore, the estimate of the lensing profiles is poor; and secondly, the emulator performs poorly, as it has been trained on predominantly lower-mass samples.

In conclusion, we have inferred the mean halo mass and linear bias for six cumulative stellar-mass threshold samples in DESI BGS Y1 using CMB lensing cross-correlations. In the following sections, we will use these best-fit linear biases to determine the effective optical depth of each sample. This is an important step that enables breaking the optical depth degeneracy. We note that using the cross-correlation with CMB alone provides weak constraints on the HOD parameters. Those can be constrained more tightly if one includes the auto-correlation of the galaxies.

\subsection{Mass dependence}
\label{sec:mass_evolution}

We first investigate the dependence of the pairwise kSZ signal on galaxy stellar mass. To this end, we divide the BGS sample into six cumulative stellar-mass thresholds: \(\log_{10}(M_\star / M_\odot) > 10, \ 10.25, \ 10.5, \ 10.75, \ 11,\) and \(11.25\). For each subsample, we measure the pairwise kSZ signal using the fiducial aperture radius \(\theta_{\rm ap} = 2.7'\) and fit the measured pairwise velocity correlation function with the theoretical model described in Section~\ref{sec:model}, allowing a single free parameter: the mean optical depth \(\bar{\tau}\). Note that in this section, we assume fiducial \textit{Planck} 2018 cosmology and linear bias provided from Table~\ref{tab:bgs_bias_mass} (the error on the bias is negligibly small compared with the error on the amplitude of the kSZ pairwise signal). In the future, when we have more constraining power, we will place joint constraints on the relevant cosmological parameters. 

The signal-to-noise ratio (SNR) is defined via the relative difference between the null model\footnote{The null model corresponds to zero pairwise kSZ signal.} and the best-fit model as:
\begin{equation}
    {\rm SNR} \equiv \sqrt{\chi^2_{\rm null} - \chi^2_{\rm bf}}
\end{equation}
where \(\chi^2_{\rm bf}\) is the best-fit \(\chi^2\) for a free amplitude, and \(\chi^2_{\rm null}\) assumes zero amplitude. In the case of one free parameter, the value of SNR gives us the significance with which that free parameter is measured for our chosen model.

We also report the probability-to-exceed (PTE) for the best-fit model, with $\chi^2 = \chi^2_{\rm bf}$, giving us the probability of obtaining a higher $\chi^2$ value than $\chi^2_{\rm bf}$, defined as:
\begin{equation}
    {\rm PTE} = \int_{\chi^2_{\rm bf}}^\infty \chi^2_{\rm m}(x) \ dx
\end{equation}
where $\chi^2_{\rm m}$ is the $\chi^2$ distribution for $m$ degrees of freedom. Unlikely events, or those in tension with theory given the experimental uncertainties, have a low PTE. Consistently high PTEs might imply experimental uncertainties have been overestimated.

We note that this approach for estimating the PTE neglects correlations in the covariance matrix $C$, which distort the true distribution of $\chi^2 = (d - m)^{\top} C^{-1} (d - m)$, where $d$ and $m$ are the data and model, respectively. To obtain an accurate estimate of the PTE we instead generate Gaussian realizations of the data vector $x \sim \mathcal{N}(0, C)$, and compute their quadratic forms
\begin{equation}
    \chi^2_{\rm mc} = x^{\top} C^{-1} x,
\end{equation}
which yield the correct null distribution of $\chi^2$ given the full covariance structure. The PTE is then defined as the fraction of Monte Carlo samples with $\chi^2_{\rm mc} > \chi^2_{\rm bf}$. This approach properly accounts for correlations in the data covariance and provides a robust measure of goodness-of-fit. We find that for all cases considered here, the difference between the simplified PTE and the MC one is less than 2\%.

\begin{figure}[ht]
    \centering
    \includegraphics[width=0.48\textwidth]{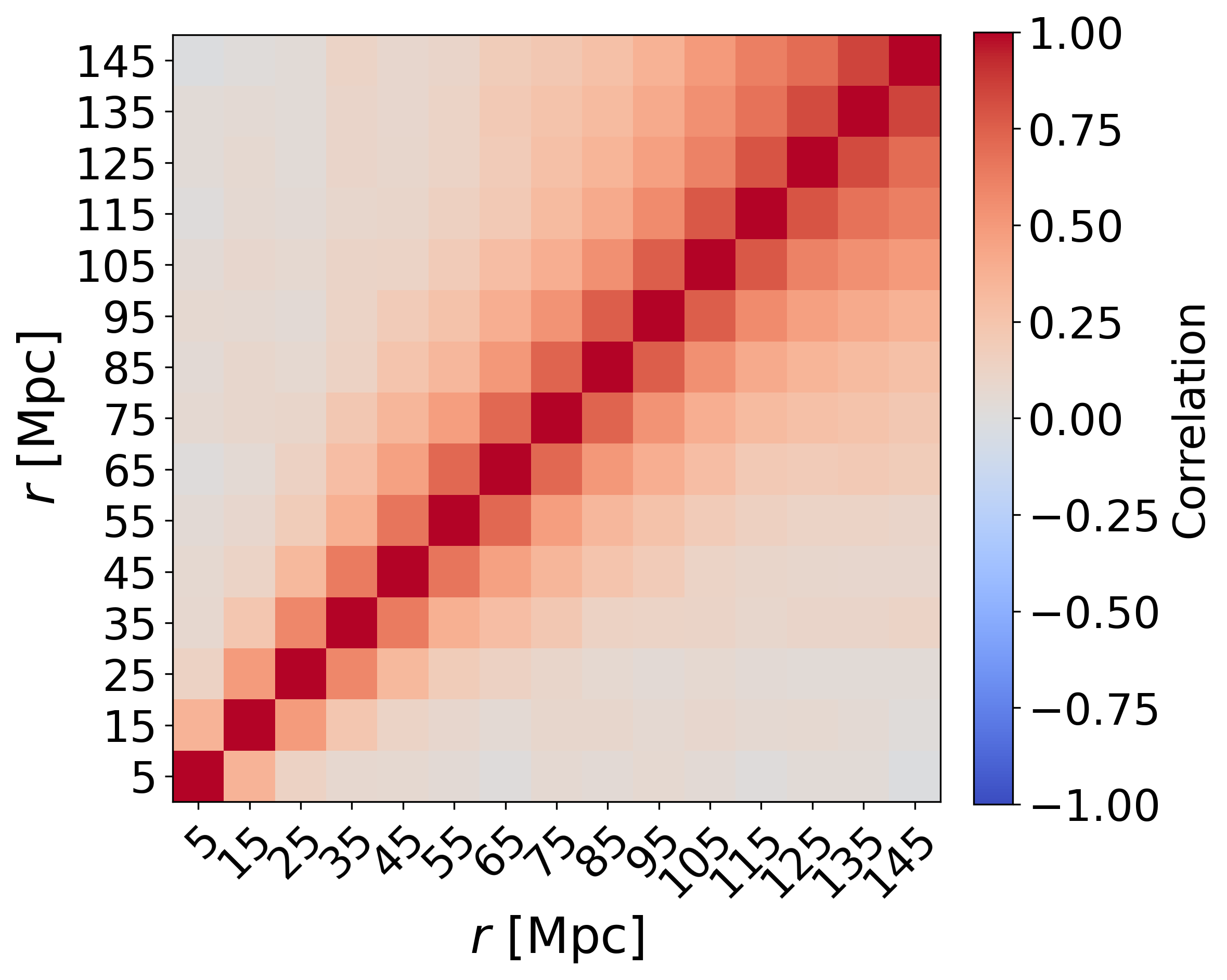}
    \caption{Correlation matrix of the pairwise estimator for the fiducial choice  \(\theta_{\rm ap} = 2.7'\) and \(\log_{10} M_\star/M_\odot > 11\), obtained from 1000 bootstrap realizations of the temperature decrements with fixed positions. The dominant noise source is the CMB temperature map. Off-diagonal correlations increase at large separations, as expected from sample variance and survey geometry, while small scales remain nearly uncorrelated due to Poisson statistics. The condition number is $\sim 30$, indicating a well-behaved, invertible matrix.}
    \label{fig:cov}
\end{figure}

In Fig.~\ref{fig:cov}, we show the correlation coefficient matrix for the fiducial choice \(\theta_{\rm ap} = 2.7'\) and 
\(\log_{10} M_\star/M_\odot > 11\), computed via bootstrap resampling of the temperature decrements holding the positions fixed, in 1000 realizations, as the dominant source of noise is the CMB temperature map. We refer the reader to Refs.~\citep{2021PhRvD.104d3502C} and \citep{2016MNRAS.461.3172S}, where the authors examine a number of different ways for computing the covariance matrix and find satisfactory results by performing bootstraps at the catalog level. 
The matrix exhibits some off-diagonal covariance between nearby radial bins, which increases slightly at large scales. This is typical of configuration-space covariance matrices, as the small scales are dominated by Poisson statistics, and thus have relatively uncorrelated errors, while the large scales are dominated by sample variance of long-wavelength modes and survey geometry, which couples the spatial information. The condition number of the covariance is $\sim 30$, indicating a well-behaved, invertible matrix.

\begin{figure*}
    \centering
    \includegraphics[width=0.48\linewidth]{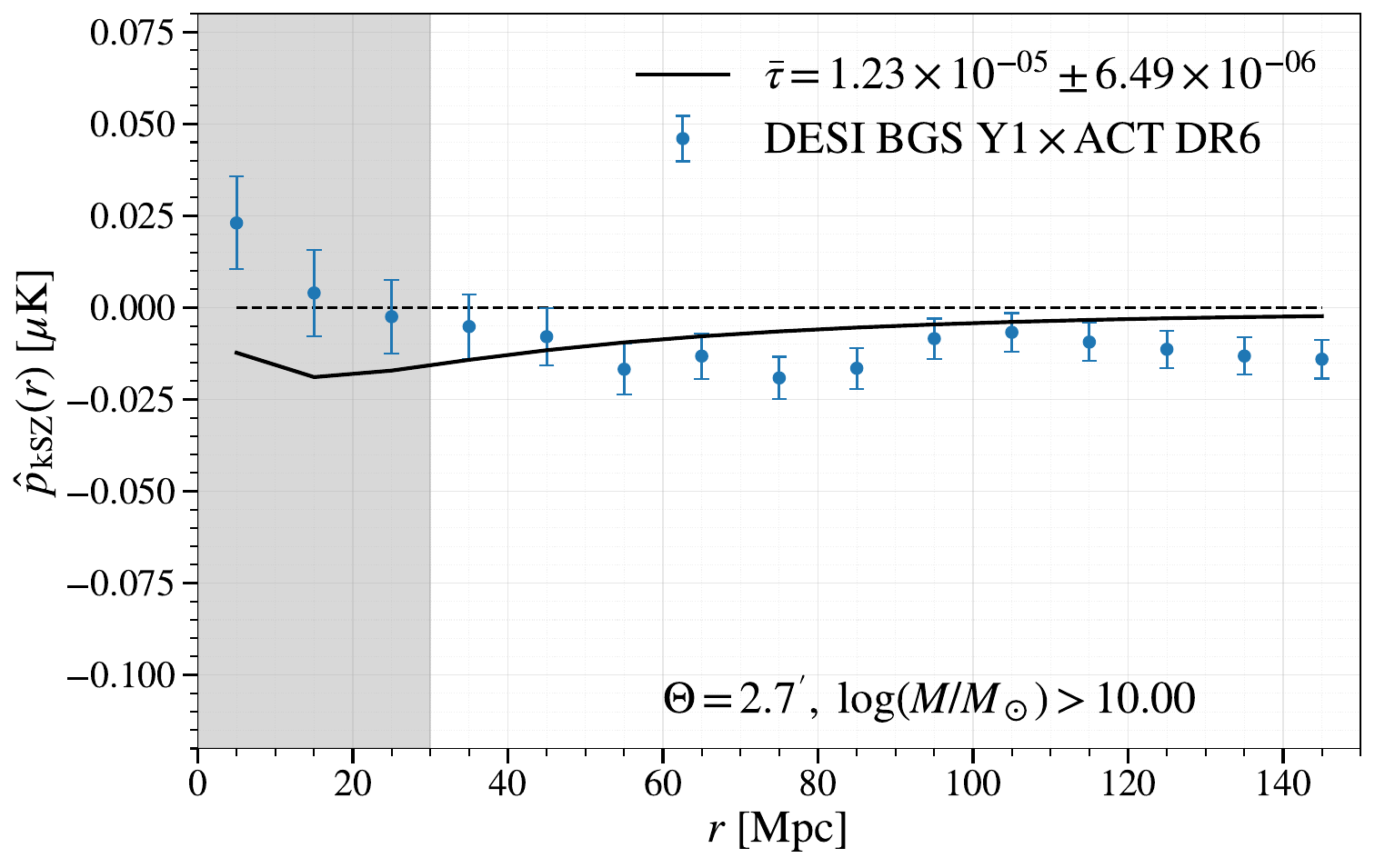}
    \includegraphics[width=0.48\linewidth]{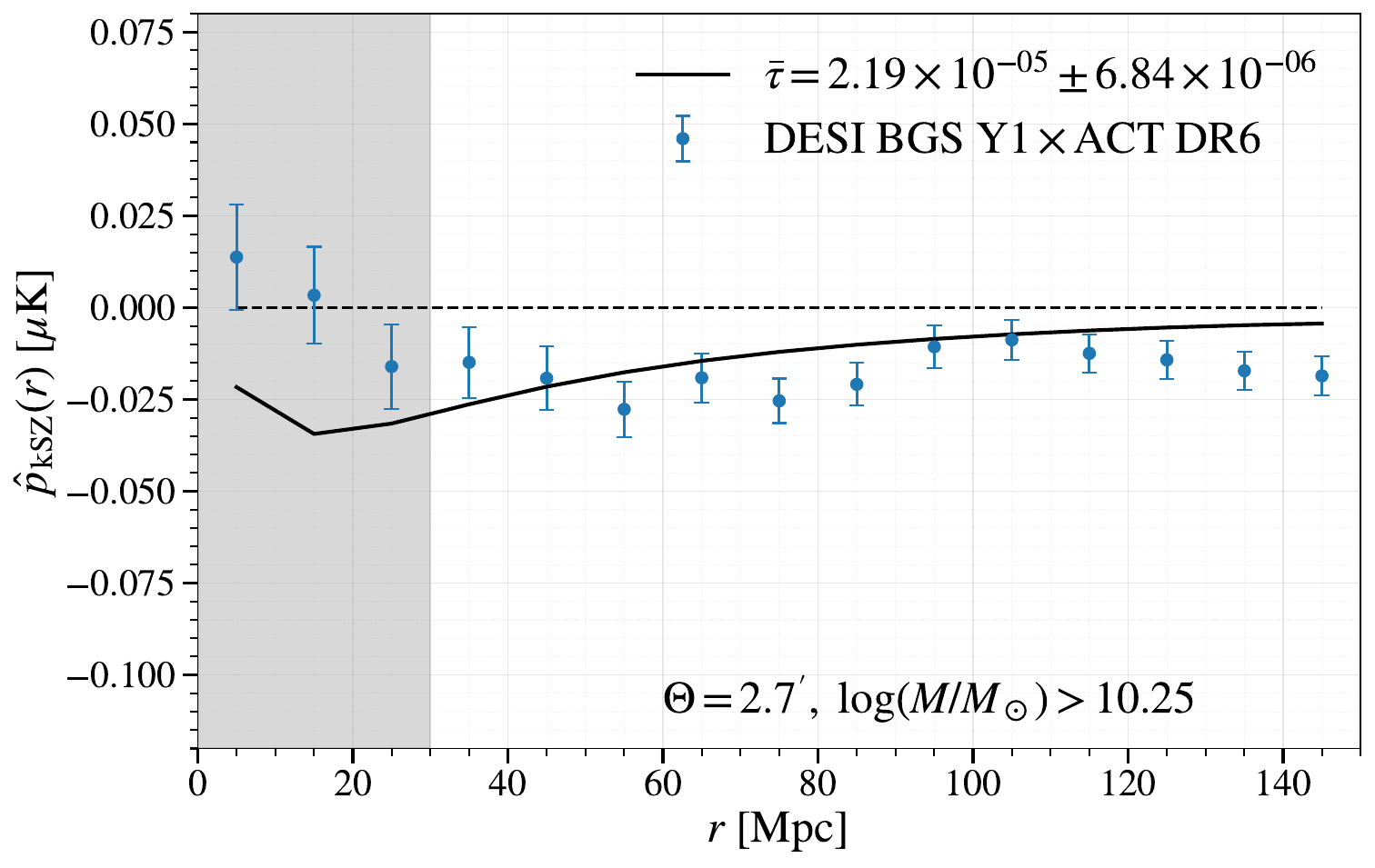}\\
    \includegraphics[width=0.48\linewidth]{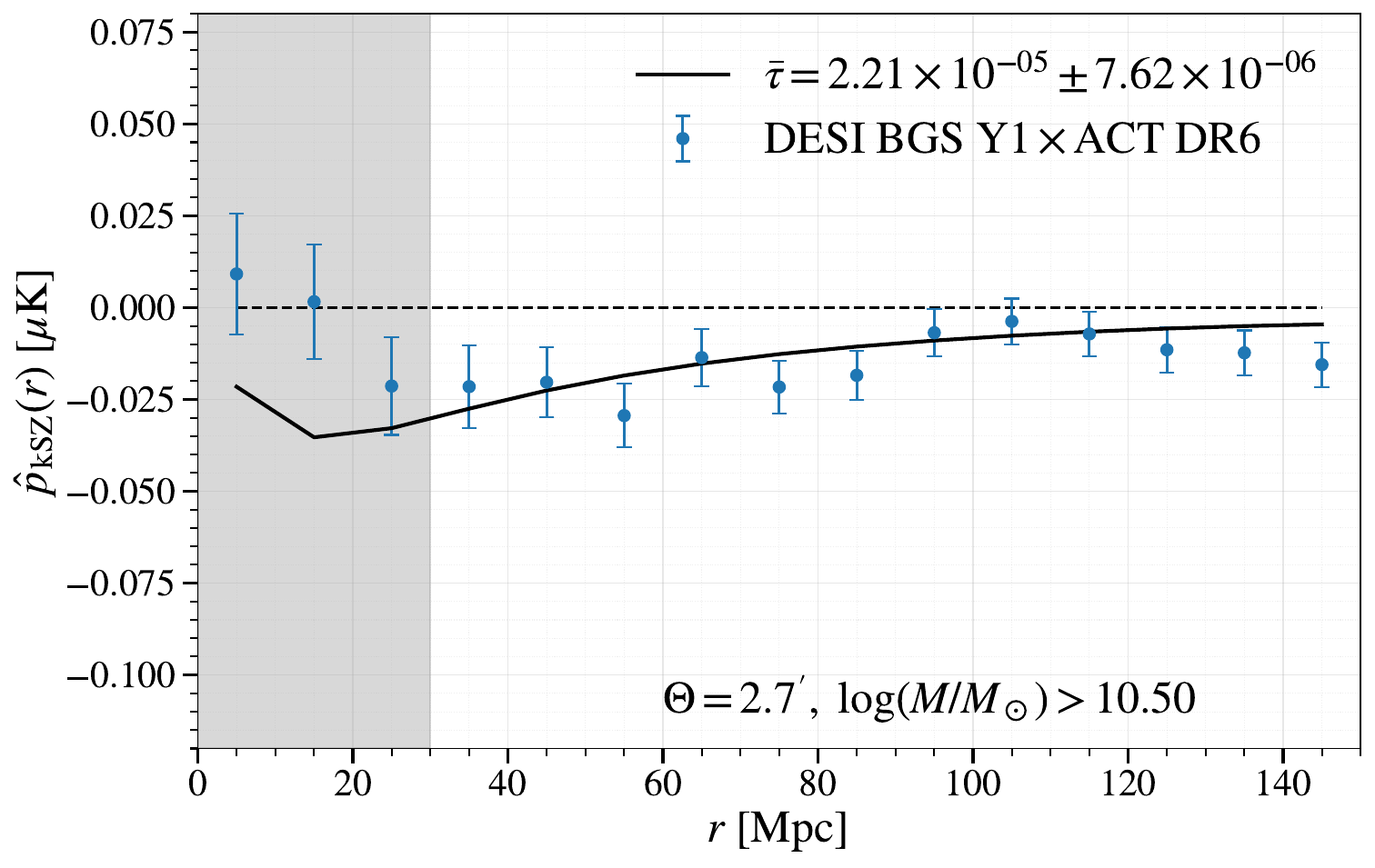}
    \includegraphics[width=0.48\linewidth]{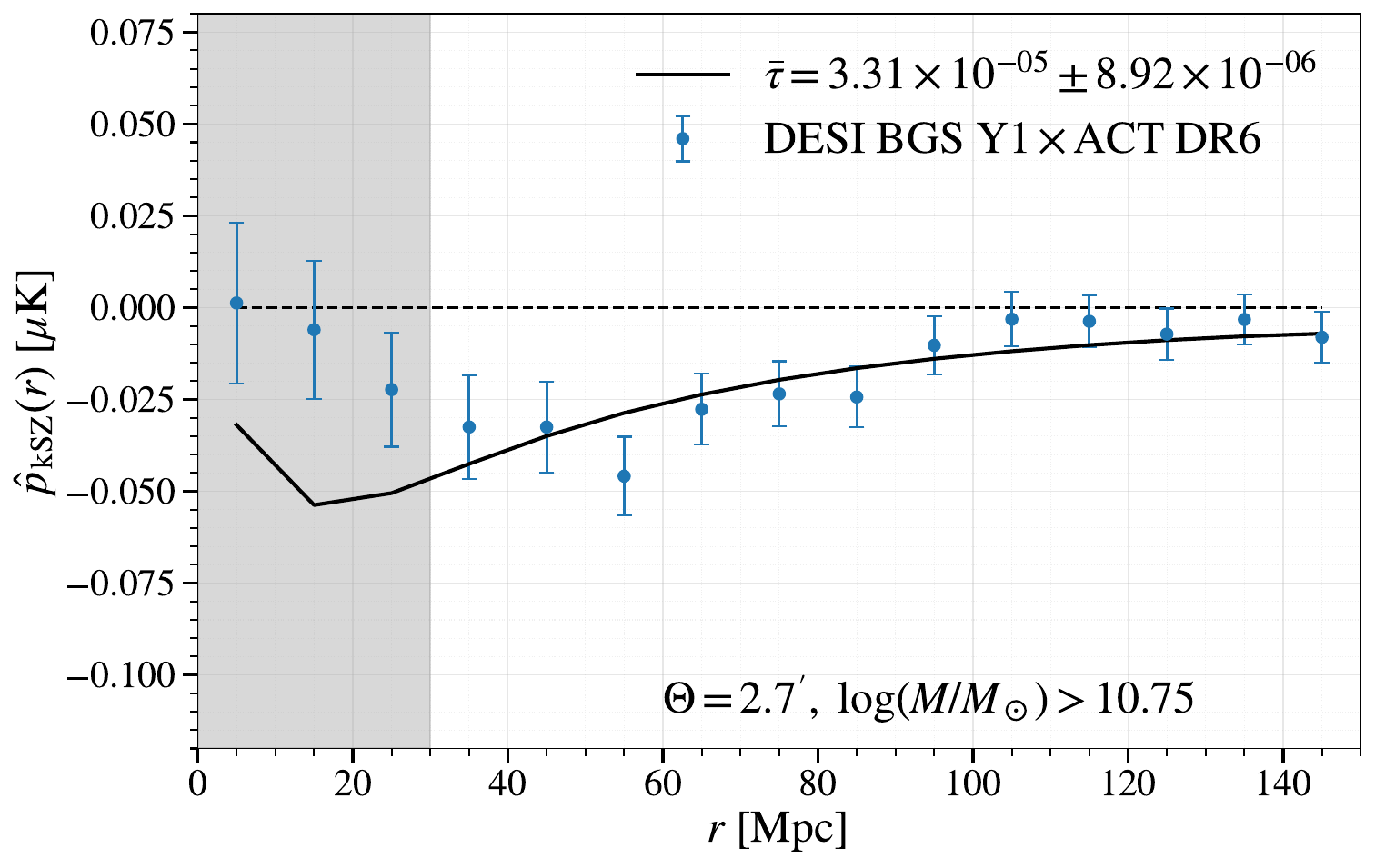}\\
    \includegraphics[width=0.48\linewidth]{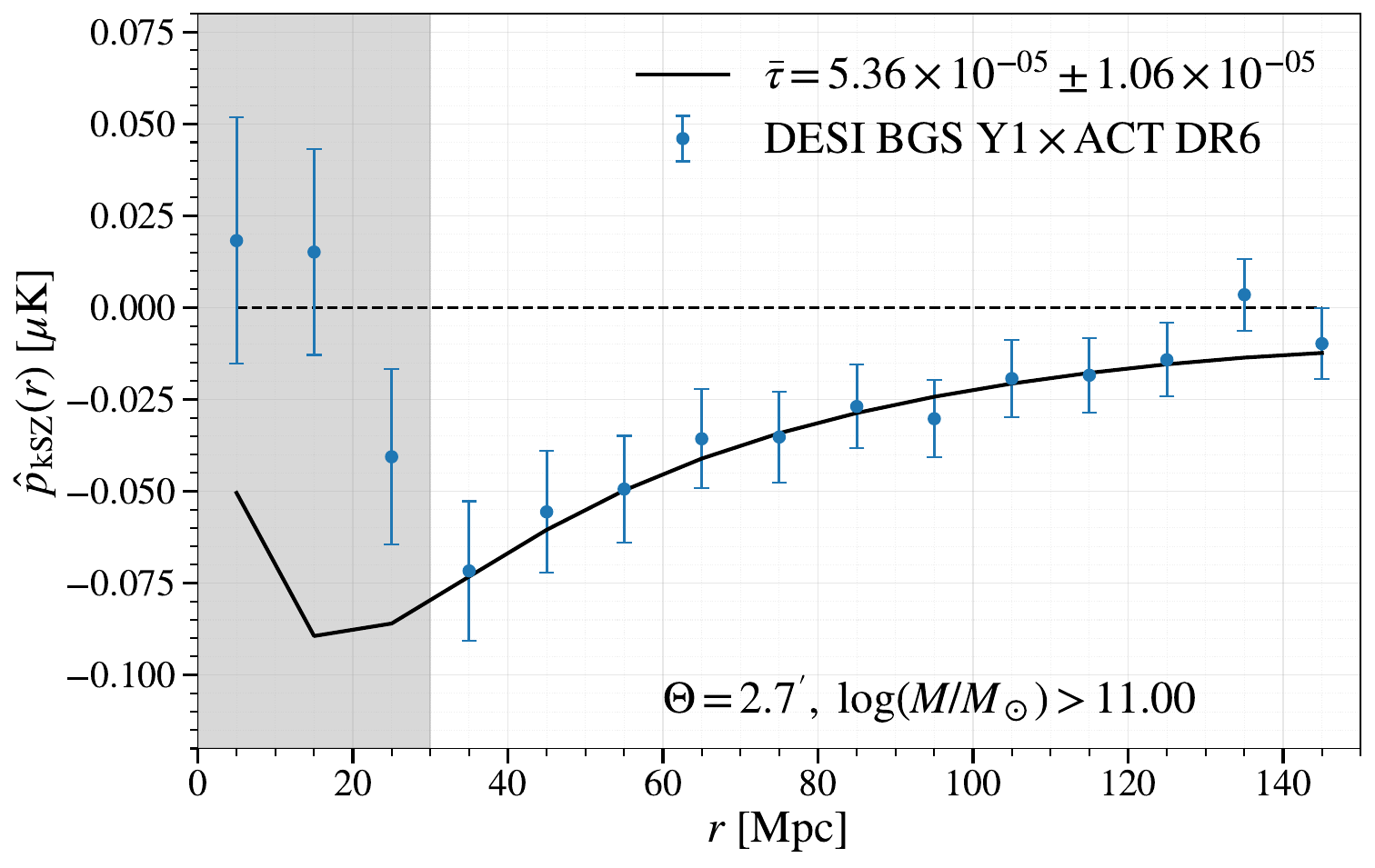}
    \includegraphics[width=0.48\linewidth]{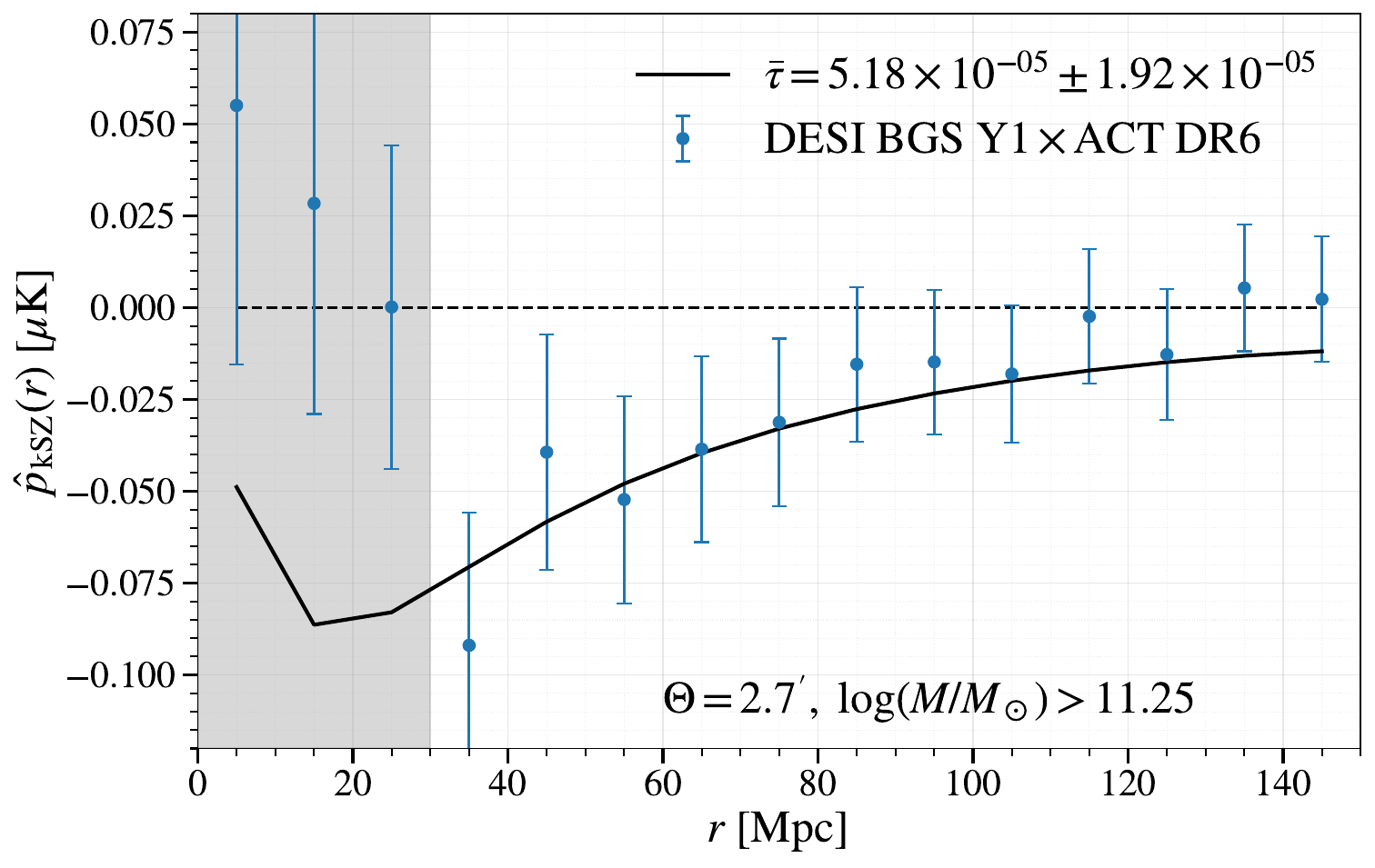}\\
    \caption{
    Pairwise kSZ measurements for the BGS sample split into stellar-mass thresholds, measured with a fixed aperture of \(\theta_{\rm ap} = 2.7'\).  
    Each curve shows the measured pairwise velocity correlation function \(\hat{p}_{\rm kSZ}(r)\) with 1$\sigma$ bootstrap uncertainties (error bars) and the best-fit model (solid lines) for the corresponding mass threshold.  
    The SNR is computed relative to the null model as described in the text.  
    The kSZ amplitude increases with stellar mass, but the SNR peaks for the \(\log_{10} M_\star > 11\) sample due to the balance between increasing halo mass and decreasing number of galaxy pairs. In this work, we focus on scales $r > 30 \ {\rm Mpc}$ where the theory is expected to be robust (see Fig.~\ref{fig:pairwise_theory}), and thus shade the $r < 30 \ {\rm Mpc}$ region.
    }
    \label{fig:mass_evolution}
\end{figure*}

In Table~\ref{tab:mass} and Fig.~\ref{fig:mass_evolution}, we show the pairwise kSZ measurements for the different stellar-mass thresholds. We find a clear mass dependence in the amplitude of the kSZ signal: the best-fit optical depth \(\bar{\tau}\) increases with stellar-mass threshold, reflecting the scaling \(\tau \propto M_{\rm halo}\).  
However, the SNR does not increase monotonically. It rises from \({\rm SNR} \approx 2\) for the \(\log_{10} M_\star > 10.0\) sample to a maximum of \(\approx 5\) for the \(\log_{10} M_\star > 11\) bin, and then decreases for the highest-mass bin (\(\log_{10} M_\star > 11.25\)) due to the reduced number of galaxy pairs. This trade-off between increasing halo mass and decreasing galaxy number counts is expected: while more massive halos yield larger \(\tau\), fewer galaxies increase measurement noise. This is also reflected in the PTE, where we see that the lowest-mass samples exhibit much poorer fits (PTE$\sim$0.01-0.1) due to the low SNR.

\begin{table}[h!]
\centering
\begin{tabular}{ccccccc}
\hline
$\log(M_\star / M_\odot)$ & $N_{\rm gal}$ & $\bar{\tau}\ [10^{-5}]$ & SNR & $\chi^2_{\rm null}$ & dof & PTE \\
\hline
10.00 & 1,610,381 & $1.23 \pm 0.65$ & 1.90 & 22.12 & 12 & 0.10 \\
10.25 & 1,379,230 & $2.19 \pm 0.68$ & 3.20 & 36.88 & 12 & 0.01 \\
10.50 & 1,084,339 & $2.21 \pm 0.76$ & 2.89 & 31.75 & 12 & 0.05 \\
10.75 &   744,562 & $3.31 \pm 0.89$ & 3.71 & 29.65 & 12 & 0.20 \\
11.00 &   421,150 & $5.36 \pm 1.06$ & 5.08 & 38.73 & 12 & 0.37 \\
11.25 &   178,206 & $5.18 \pm 1.92$ & 2.70 & 11.75 & 12 & 0.97 \\
\hline
\end{tabular}
\caption{Pairwise kSZ measurements for BGS subsamples split by stellar-mass threshold at $\theta_{\rm ap} = 2.7'$.  
The table lists the number of galaxies, best-fit optical depth \(\bar{\tau}\), detection significance (SNR) relative to the null, and the $\chi^2_{\rm null}$ values.  
We use 12 data points and fit for 1 free parameter in each case. We see that the fit is not very good for the low-mass samples, and their SNR is below 3$\sigma$.}
\label{tab:mass}
\end{table}

\subsection{Aperture variation}
\label{sec:aperture_evolution}

We next explore the dependence of the measured kSZ signal on the aperture radius \(\theta_{\rm ap}\) used when measuring the temperature decrements around each galaxy (see Section~\ref{sec:est}). One might naively want to select a larger aperture that captures more of the gas signal and also avoids the ACT beam of 1.6 arcmin. However, while smaller apertures exclude part of the extended gas distribution around halos, reducing the measured signal, larger apertures have a much larger contamination from the primary CMB anisotropies, and the primary CMB is much larger than our signal on scales beyond $\sim$3 arcmin. Here, we focus on the \(\log_{10} M_\star > 11\) subsample, which has the highest SNR and test aperture radii of \(2.1', 2.3', 2.5', 2.7', 2.9', 3.1', 3.3', \) and \(3.5'\).

As seen in Table~\ref{tab:aperture_results}, the SNR for the $log_{10}(M_{\star})>11$ peaks at \(\theta_{\rm ap} = 3.3'\), which represents a compromise between capturing the maximum amount of signal while minimizing the noise from the primary CMB fluctuations.
This choice is consistent with the typical size of BGS halos at the mean redshift of the subsample (\(z \sim 0.3\)), which have an effective virial radius of around 0.5 Mpc, corresponding to \(\sim 2\) arcmin (when we take into account the ACT beam). Due to baryonic feedback, the gas is pushed out even further than the dark matter, to about two or three times the virial radius, i.e. 4--5 arcmin \citep[see e.g.,][for the photometric BGS gas density]{2025PhRvD.111b3534H}. This is in line with previous kSZ measurements hinting at the presence of large baryonic feedback \citep[see e.g.,][]{2024arXiv240707152H,2025arXiv250319870R}. We note that the more of the gas that is enclosed within the measured aperture, the weaker the dependence on feedback.

\begin{table}[h!]
\centering
\begin{tabular}{cccccc}
\hline
$\theta_{\rm ap}$ [arcmin] & $\bar{\tau}\ [10^{-5}]$ & SNR & $\chi^2_{\rm null}$ & dof & PTE \\
\hline
2.1 & $3.15 \pm 0.93$ & 3.40 & 23.24 & 12 & 0.47 \\
2.3 & $3.85 \pm 0.97$ & 3.99 & 26.90 & 12 & 0.53 \\
2.5 & $4.66 \pm 1.06$ & 4.41 & 29.78 & 12 & 0.58 \\
2.7 & $5.36 \pm 1.06$ & 5.08 & 38.73 & 12 & 0.37 \\
2.9 & $5.99 \pm 1.16$ & 5.15 & 40.86 & 12 & 0.28 \\
3.1 & $6.56 \pm 1.29$ & 5.10 & 40.52 & 12 & 0.27 \\
3.3 & $7.19 \pm 1.37$ & 5.24 & 42.98 & 12 & 0.21 \\
3.5 & $7.45 \pm 1.45$ & 5.13 & 42.18 & 12 & 0.20 \\
\hline
\end{tabular}
\caption{Pairwise kSZ measurements for the $\log_{10}(M_\star/M_\odot) > 11$ BGS subsample as a function of aperture radius $\theta_{\rm ap}$.  
Listed are the best-fit optical depth $\bar{\tau}$ (in units of $10^{-5}$), the detection significance (SNR) relative to the null, and the $\chi^2_{\rm null}$ evaluated with the full covariance. There are 421,150 galaxies in this sample. In all cases, we find that the PTE is good, signifying that the model fits the data well. We use 12 radial data points and fit one free amplitude parameter, the mean optical depth $\bar \tau$. Note that the amplitude of the signal, according to our model (Eq.~\ref{eq:v12_halo}) depends on the product between linear bias, $\bar \tau$ and cosmology (see Eq.~\ref{eq:taubfs82}). Here we have fixed the cosmology using the {\it Planck} 2018 best-fit parameters and the linear bias, using the CMB lensing $\kappa$ fits (see Table~\ref{tab:bgs_bias_mass}), which is known at the 5\% level.}
\label{tab:aperture_results}
\end{table}

The SNR trend shows that apertures smaller than the virial scale miss part of the kSZ signal, while larger apertures dilute the signal with additional noise.
The fiducial aperture of \(2.7'\) captures most of the kSZ contribution from BGS halos, making it an optimal choice for the full sample. We note that for higher-mass halos, the optimal value of the aperture will be larger.
The fractional error on \(\bar{\tau}\) increases slightly for the largest apertures, reflecting the increased uncertainty from the primary CMB contamination.

\begin{figure*}
    \centering
    \includegraphics[width=0.48\linewidth]{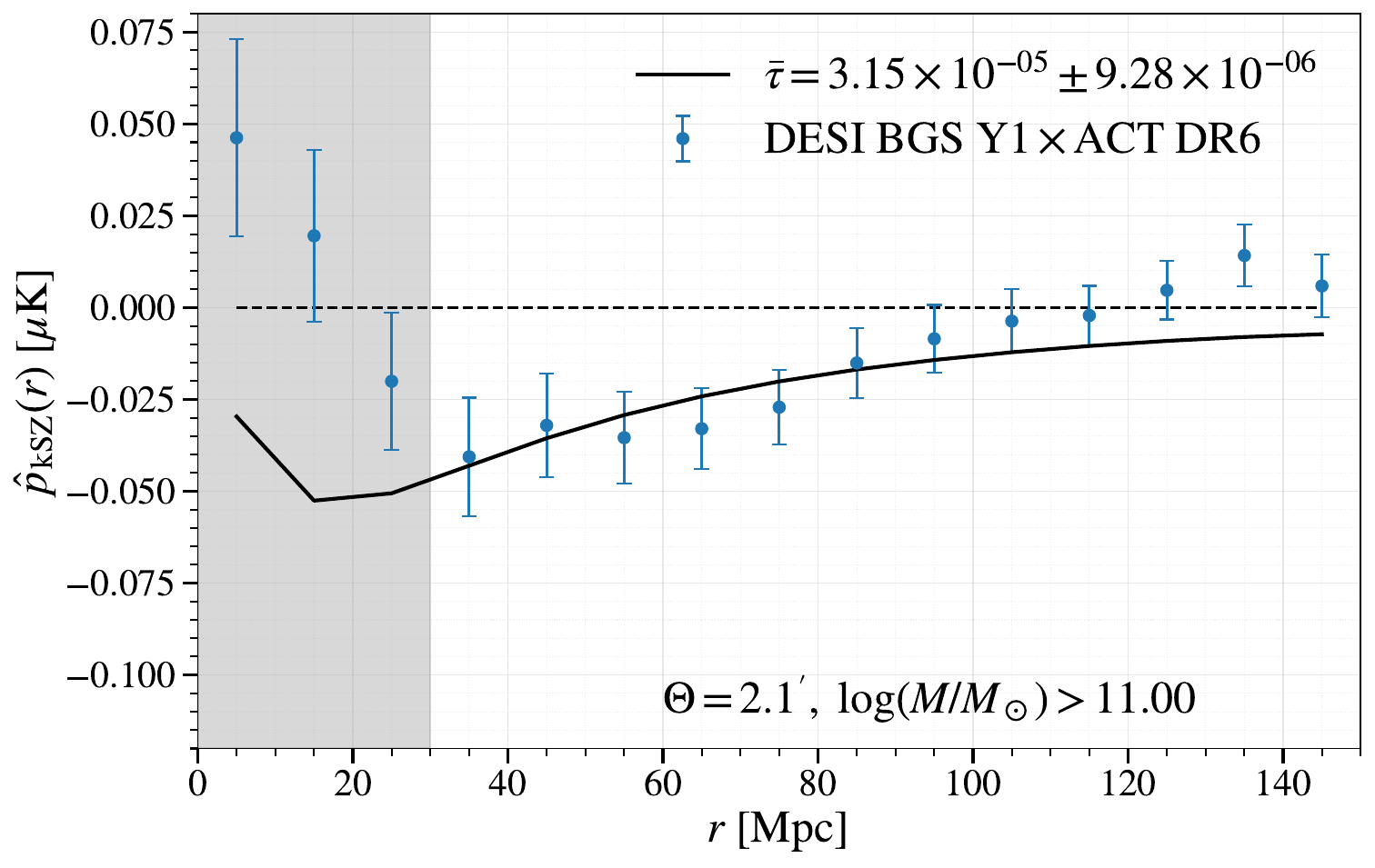}
    \includegraphics[width=0.48\linewidth]{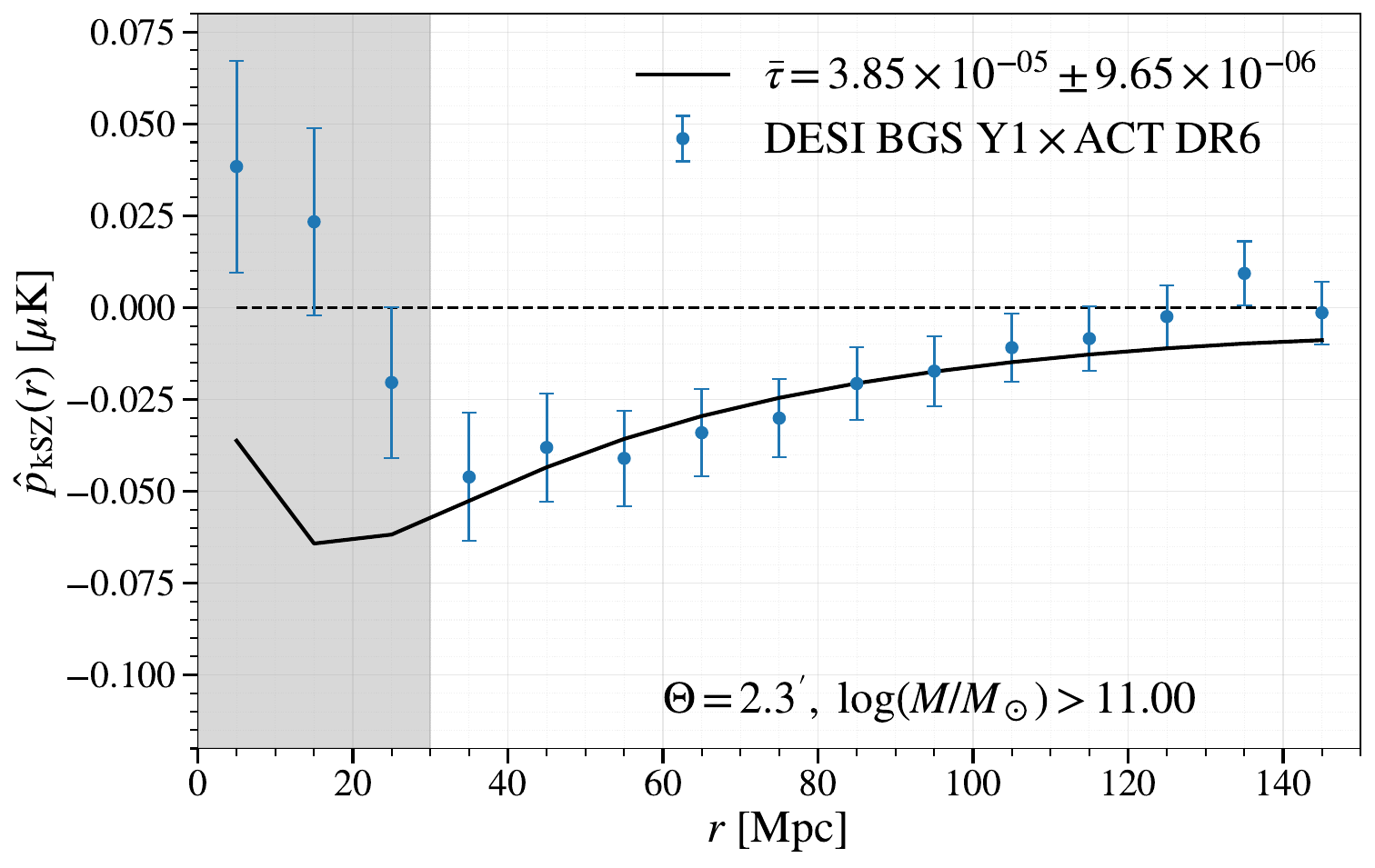}\\
    \includegraphics[width=0.48\linewidth]{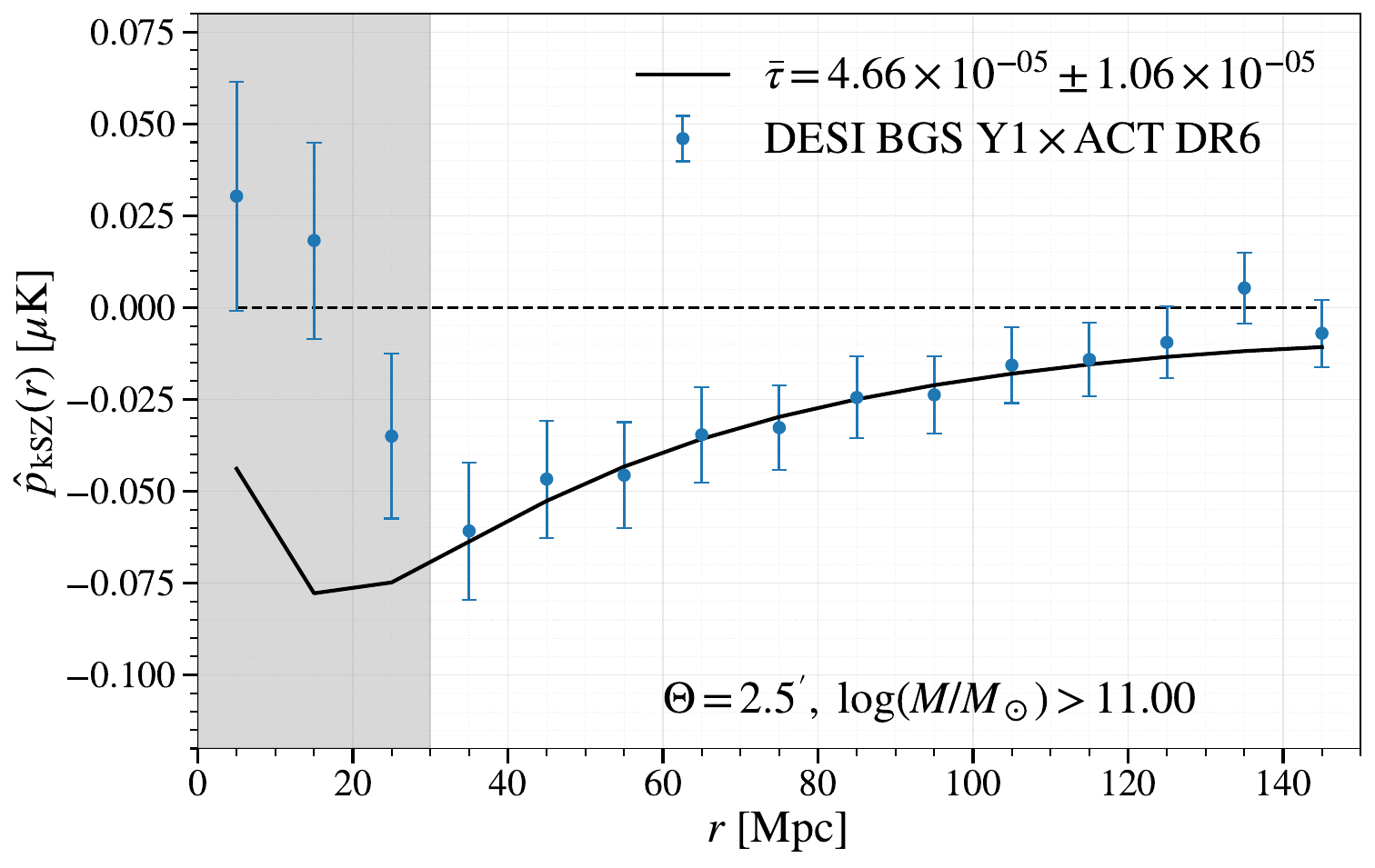}
    \includegraphics[width=0.48\linewidth]{figs/BGS_ACT_DR6_fixedTh2.70_logm11.00_PV_boot.pdf}\\
    \includegraphics[width=0.48\linewidth]{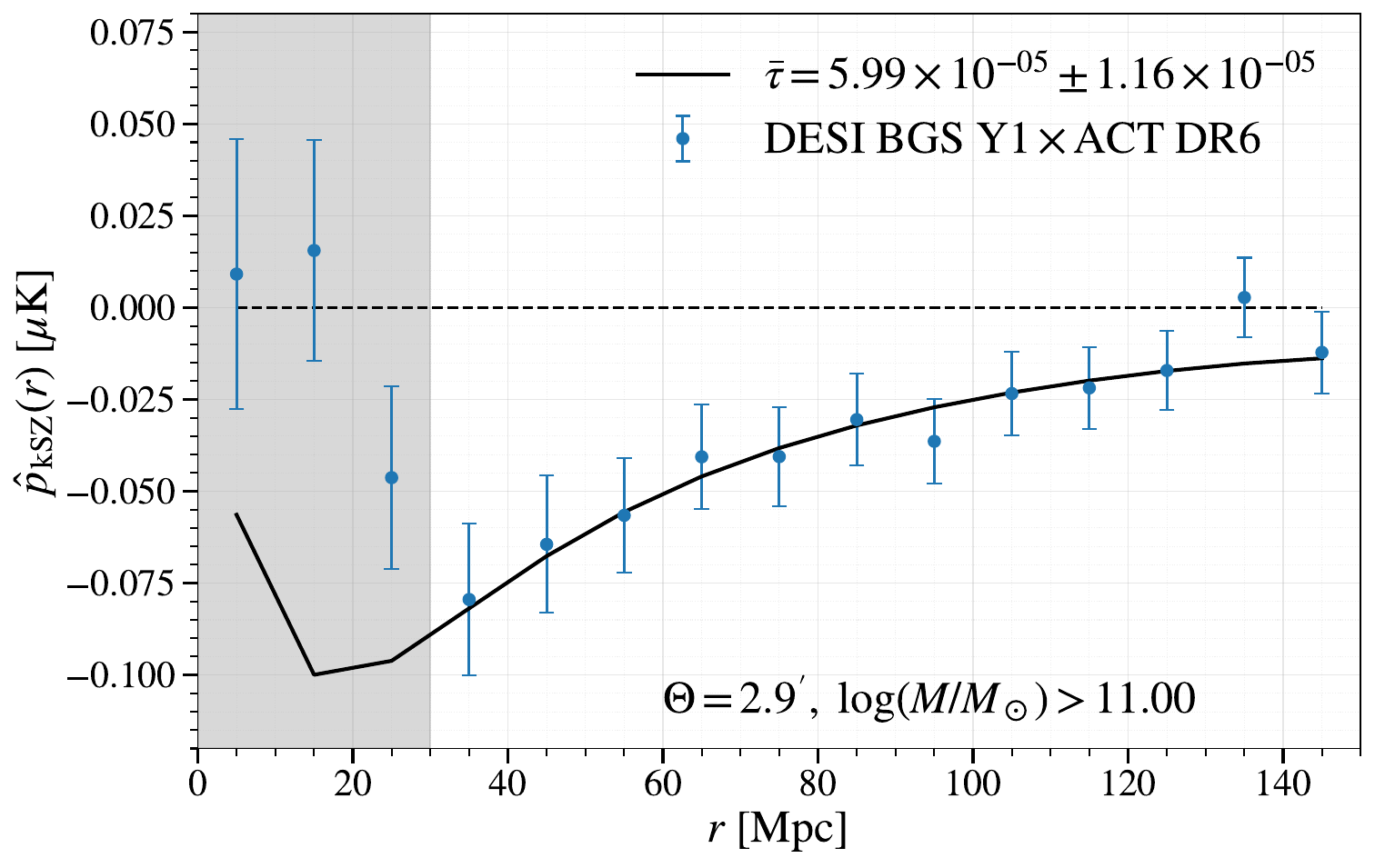}
    \includegraphics[width=0.48\linewidth]{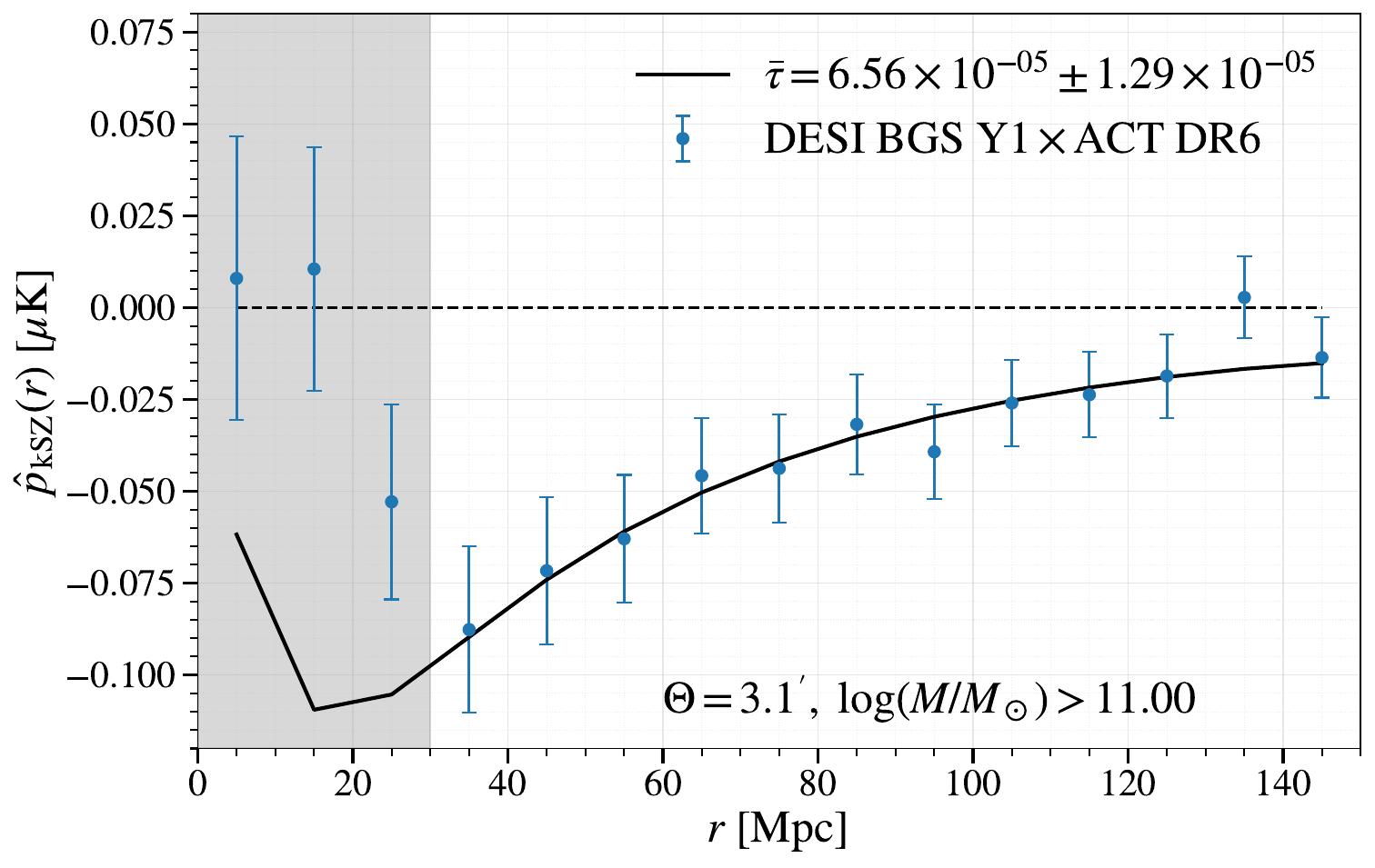}\\
    \caption{
    Dependence of the pairwise kSZ SNR on aperture radius \(\theta_{\rm ap}\) for the \(\log_{10} M_\star > 11\) sample.  
    The SNR peaks at the fiducial \(2.7'\), balancing the inclusion of the majority of the kSZ signal from the galaxy’s gaseous halo with the suppression of contamination from primary CMB fluctuations and instrumental noise.  
    The table lists the corresponding best-fit optical depths \(\bar{\tau}\) and the fraction of the signal constrained by the measurement. In this work, we focus on scales $r > 30 \ {\rm Mpc}$ where the theory is expected to be robust (see Fig.~\ref{fig:pairwise_theory}), and thus shade the $r < 30 \ {\rm Mpc}$ region.
    }
    \label{fig:aperture_evolution}
\end{figure*}

\subsection{Null tests}
\label{sec:null_tests}

To confirm that the measured kSZ signal is not driven by systematics, we perform two null tests:  
(1) Randomly shuffling the signs of the pairwise kSZ temperature differences;  
(2) Randomly shuffling the sky positions of the galaxies.  
Both procedures destroy any coherent pairwise signal while preserving the statistical properties of the noise, allowing us to verify that our bootstrap covariance correctly captures the uncertainties and that contamination from other sources (e.g., CIB, tSZ) is negligible.

For each test, we compute the \(\chi^2\) of the measured \(\hat{p}_{\rm kSZ}(r)\) relative to zero of the $log_{10}(M_{\star})>11$ sample and report the probability-to-exceed (PTE).  
The results are summarized in Table~\ref{tab:null_tests}.  
Both null tests are consistent with zero signal, confirming that the detection in the main analysis is not due to spurious correlations.

\begin{table}[h!]
\centering
\begin{tabular}{lccccc}
\hline
Null test & $\bar{\tau} \ [10^{-6}]$ & SNR & $\chi^2_{\rm null}$ & dof & PTE \\
\hline
Random sign      & $-1.30 \pm 6.25$ & 0.21 & 10.08 & 12 & 0.61 \\
Random position  & $-1.67 \pm 10.5 $ & 0.16 &  8.56 & 12 & 0.74 \\
\hline
\end{tabular}
\caption{Null test results for the \(\log_{10} M_\star > 11\) BGS subsample at the fiducial aperture \(\theta_{\rm ap} = 2.7'\).  
We report the best-fit optical depth \(\bar{\tau}\), SNR relative to zero, \(\chi^2_{\rm null}\), degrees of freedom (dof), and probability-to-exceed (PTE). The PTE values suggest that the signal is consistent with the null model, as expected.}
\label{tab:null_tests}
\end{table}

\begin{figure*}
    \centering
    \includegraphics[width=0.48\linewidth]{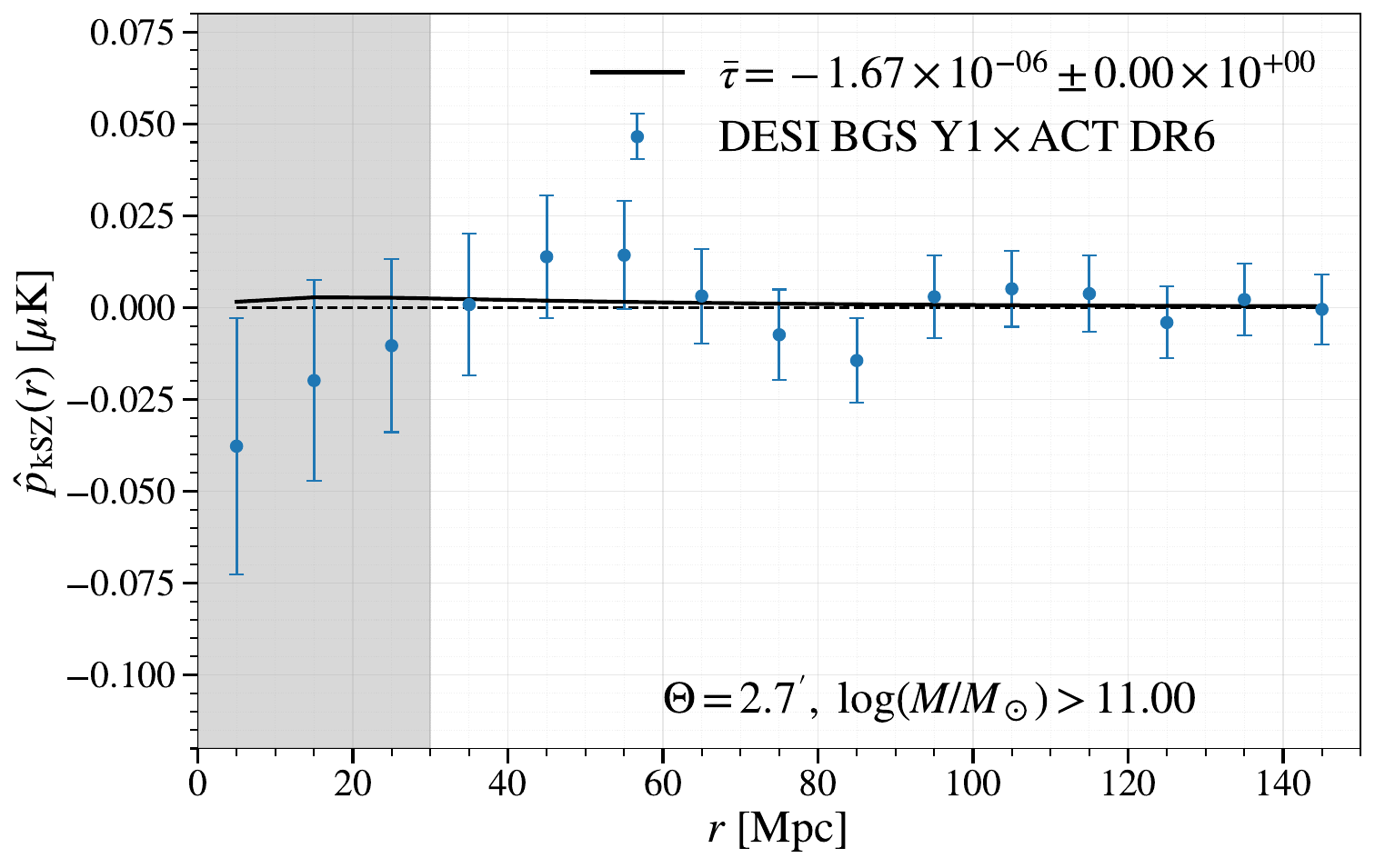}
    \includegraphics[width=0.48\linewidth]{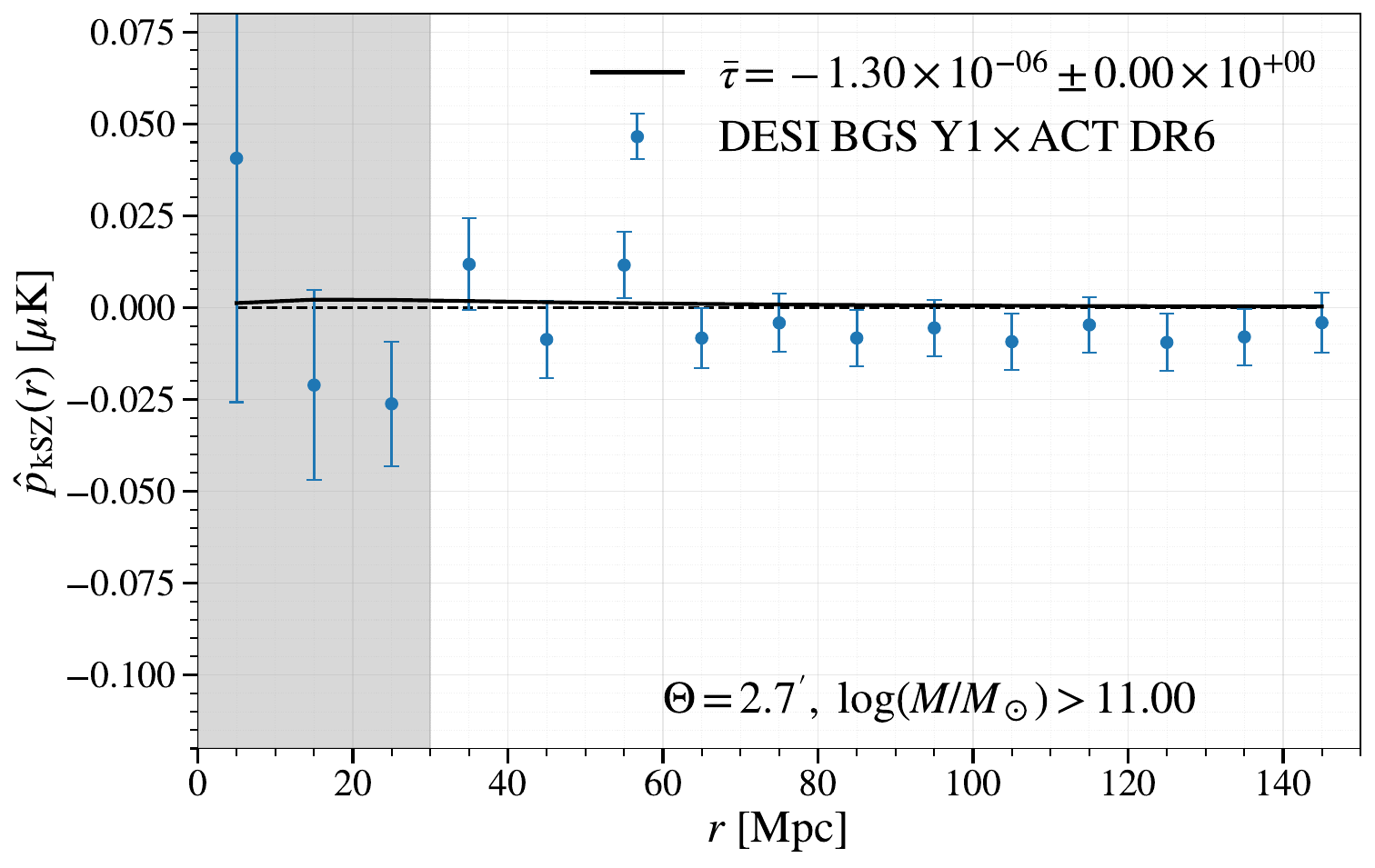}
    \caption{
    Null test results for the \(\log_{10} M_\star > 11\) BGS subsample at the fiducial aperture \(\theta_{\rm ap} = 2.7'\).  
    \textit{Left:} random sign shuffling of the pairwise kSZ temperature differences.  
    \textit{Right:} random shuffling of galaxy sky positions.  
    In both cases, the measured signal is consistent with zero, with \(\chi^2_{\rm null}\) and PTE values reported in Table~\ref{tab:null_tests}, demonstrating the robustness of our main detection.  
    Error bars are estimated via jackknife resampling.
    }
    \label{fig:null_tests}
\end{figure*}

\subsection{Cosmological interpretation}
\label{sec:cosmo_constraints}

The pairwise kSZ signal is particularly sensitive to the velocity correlation function, providing a direct probe of the dynamics of the Universe.  
However, there exists an \emph{optical depth degeneracy}: the measured signal constrains only the product of \(\bar{\tau}\) and cosmological parameters such as \(f\sigma_8^2 H_0\).  
Breaking this degeneracy requires accurate estimates of \(\bar{\tau}\), which can be done in various ways. For example, one can estimate $\bar \tau$ from hydrodynamical simulations that reproduce the observed galaxy sample properties and mean host halo masses.  
This approach is challenging because it depends on well-constrained halo occupation distributions (HODs) to match the galaxy population precisely as well as having a trustworthy subgrid model for feedback in the simulation.  Other ways of inferring the optical depth include using the tSZ flux of the same sources and taking advantage of the relatively tight $Y$--$\tau$ relationship \citep{2021PhRvD.104d3503V}, estimating $\bar{\tau}$ from patchy screening measurements \citep{2024arXiv240113033C,2025arXiv250617217S,2025arXiv250617379H}, and semi-analytical approaches linking total mass, thermal energy and gas mass.

An appropriate hydrodynamical simulation for calibrating the optical depth measure should (i) implement a feedback model that captures the baryonic processes shaping the intra-halo medium, (ii) have sufficient volume to sample the high-mass end of the halo mass function, and (iii) resolve the relevant gas distributions on the scales probed by the kSZ signal. While different simulations vary in their feedback prescriptions and resolutions, our choice of \textsc{SIMBA} reflects a balance of these criteria and its closer agreement with current observational constraints. 

We chose to use the \textsc{SIMBA} simulation \citep{2019MNRAS.486.2827D} for this task for the following reason. Recent analyses of stacked kSZ measurements \citep[e.g.][]{2024arXiv240707152H,2024MNRAS.534..655B,2025arXiv250319870R,2025MNRAS.540..143M} indicate that the data prefer stronger feedback than is implemented in, for instance, the IllustrisTNG simulation. In addition, a sufficiently large simulation volume is necessary to robustly sample the most massive halos that dominate the signal. SIMBA provides both a comparatively strong feedback model and a large cosmological volume, making it particularly suitable for this proof-of-concept calibration. Its assumed cosmology is also concordant with \textit{Planck} 2018.

Here, we explore the cosmological implications of our measurement by placing a prior on the mean optical depth \(\bar{\tau}\), derived from hydrodynamical simulations, in order to break the optical depth degeneracy.  
As already stated, the \textsc{SIMBA} simulation includes relatively stronger AGN feedback than other hydrodynamical simulations \citep{2023MNRAS.523.2247S}. In Appendix~\ref{app:simba}, we show that at 3.5 arcmin, the stacked kSZ profiles from the photometric BGS sample selected in Ref.~\citep{2025PhRvD.111b3534H} with halo mass reported in Ref.~\citep{2025arXiv250714136H}, on the one hand, and a closely matching BGS-like sample in SIMBA, on the other, agree at the 1$\sigma$ level. In particular, our simulated galaxies are selected to match the halo masses inferred from CMB lensing measurements of the BGS sample, allowing us to assign their mean optical depth \(\bar{\tau}_{\rm sim}\) in a way that is representative of our observed population.  We note that as we go to higher halo masses, where feedback has a smaller impact on the  gas distribution (the gas distribution is shaped more strongly by gravity), we expect the predictions from different hydrodynamical simulations to agree, independent of the feedback prescription. 
For this reason, in this work we will focus on the higher-mass sample, $\log M_{\ast}/M_{\odot} > 11$, for which we also attain the highest signal-to-noise. We leave a more detailed study of baryonic feedback in the BGS sample for future work.

We note that there are several caveats when comparing the signal between data and simulations at such low redshifts. Some of these issues include geometric and projection effects as well as baryonic feedback mismatch. Because the angular size of the BGS-hosting galaxy groups varies rapidly, the mean optical depth at fixed aperture actually probes a huge range of physical scales, including scales that are very close to the halo center, where the gas has potentially already been pushed out, as well as very far from the center, where we are measuring mostly noise. These effects lead to scale-dependent effects on $\bar \tau$ as a function of aperture, and the proper way to test them is via light cone simulations. Moreover, if the baryonic feedback prescription in the simulation is a poor match to the effective feedback in the Universe, that may lead to additional differences in the $\bar \tau$ evolution with aperture.
In principle, both the stacked kSZ and the pairwise kSZ should yield the same $\tau(r)$ profile. We leave a more detailed comparison between the gas density profiles inferred using the kSZ stacking technique and the gas density profiles inferred using the pairwise kSZ estimator to a future study.

As noted earlier in the text (see Eq.~\ref{eq:taubfs82}), the amplitude of the pairwise signal depends on the product between optical depth, \(\bar{\tau}_{\rm sim}\), and \(f\sigma_8^2\).
Thus, with \(\bar{\tau}_{\rm sim}\) as a prior, we can refit the pairwise velocity correlation function to place constraints on the growth-rate combination \(f\sigma_8^2\) at \(z = 0.33\). 
Although the resulting constraints are relatively weak and limited due to the significance of detection of the pairwise signal, as seen in Table~\ref{tab:aperture_results}, this exercise demonstrates the unique potential of the pairwise kSZ effect to probe large-scale velocity fields and test gravity on cosmological scales. 

\begin{figure*}
    \centering
    \includegraphics[width=0.98\textwidth]{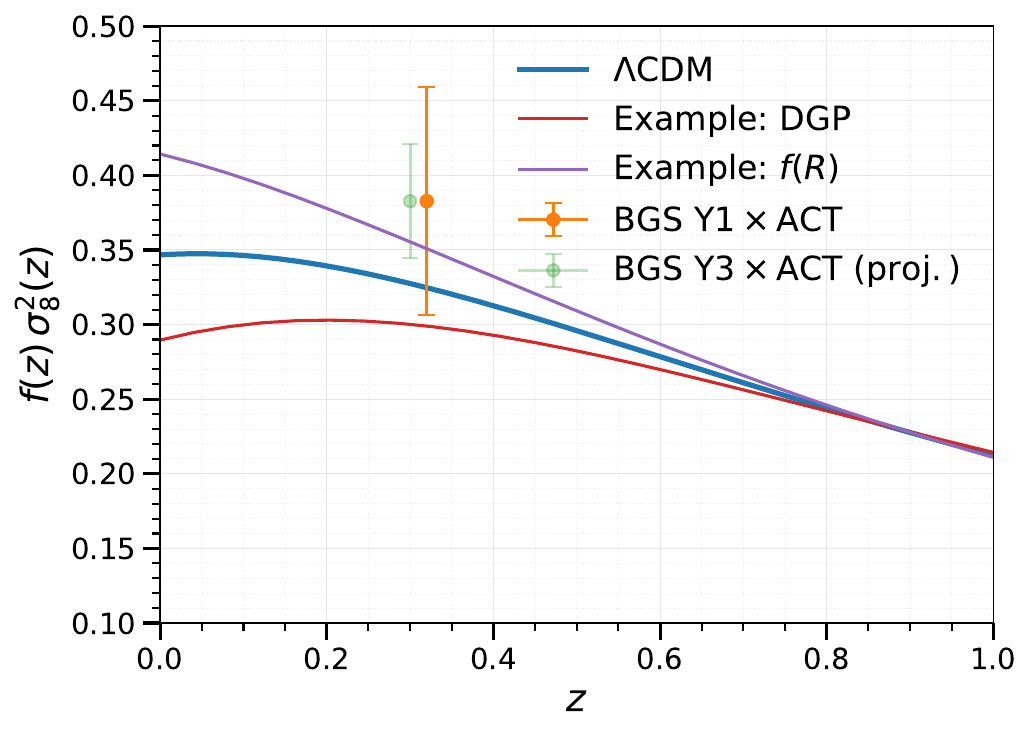}
    \caption{
    Cosmological constraints from the pairwise kSZ measurement of the \(\log_{10} M_\star > 11\) BGS sample at the fiducial aperture \(\theta_{\rm ap}=3.5'\).  
    The mean optical depth \(\bar{\tau}\) is fixed to the value inferred from the \textsc{SIMBA} simulation, yielding constraints of $f\sigma_8^2 = 0.38 \pm 0.076$ at \(z=0.33\) (the mean redshift of that sample as seen in Fig.~\ref{fig:logm}).  
    Although the current constraints are weak, this redshift range is particularly interesting for testing modified gravity models, where deviations from GR are expected to be most pronounced. 
    We also show approximate curves for DGP ($\gamma = 0.68$) and Hu--Sawicki $f(R)$ with $f_{R0} = 10^{-5}$ ($\gamma = 0.43$), obtained by replacing the GR growth index in $f(z)$ with the corresponding values and rescaling $\sigma_8(z)$ accordingly. These are only approximate and are included for illustrative purposes.
    Future data (DESI DR2) are anticipated to reduce the error bars by roughly a factor of two (as the number of objects will increase substantially), enabling meaningful constraints on alternative gravity scenarios.
    }
    \label{fig:cosmo_constraints}
\end{figure*}

We show the results from this exercise in Fig.~\ref{fig:cosmo_constraints}, finding $f\sigma_8^2 = 0.38 \pm 0.076$. 
To arrive at that number, we have made the assumption that SIMBA provides a good match for the value of the mean optical depth at 3.5 arcmin for the $\log M_{\ast}/M_{\odot} > 11$ sample, namely $\bar \tau [10^{-5}] = 6.75$. Because the simulation-inferred value of the mean optical depth
is about 10\% smaller than the empirically measured value of $\bar \tau [10^{-5}] = 7.45$ in Table~\ref{tab:aperture_results}, we obtain an \(f\sigma_8^2\) that is higher by 10\% compared with $Planck$ 2018. We emphasize again that the constraints on cosmology are completely dependent on our assumption for the optical depth. 

We describe in more details how well SIMBA fits the kSZ gas density profiles and how the sample is selected in App.~\ref{app:simba}. In short, since our sample is selected by stellar mass (and the stellar mass estimate for each galaxy has a pretty large scatter, $\gtrsim$0.1 dex), we mimic this in the simulation by rank ordering the galaxies by their stellar masses and making a cut-off at $\bar n_{\rm g} = 2 \times 10^{-3} \ ({\rm Mpc}/h)^{-3}$ to account for the large scatter in stellar mass. Then we impose a halo mass threshold of $\log (M_{\rm h}/M_\odot/h) < 14.7$ to mimic the cluster mask applied to our ACT DR6 temperature map when performing the kSZ pairwise measurement. The resulting mean halo mass of the simulation sample is $\langle \log (M_{\rm h}/M_\odot/h) \rangle \approx 13.38$, which matches very closely the mean halo mass of that same sample in the data, as inferred from CMB lensing (see Table~\ref{tab:bgs_bias_mass}).

For comparison, we show approximate predictions for two modified gravity scenarios. In particular, we adopt the Dvali–Gabadadze–Porrati (DGP) model, where the growth rate is well described by $f(z) \approx \Omega_m(z)^{\gamma}$ with $\gamma \simeq 0.68$, and the Hu--Sawicki $f(R)$ model with $f_{R0} = 10^{-5}$, for which the growth index can be approximated as $\gamma \simeq 0.43$ \citep[see][for a recent review]{2022GReGr..54...44S}. In both cases, we compute $\sigma_8(z)$ by rescaling the $\Lambda$CDM $\sigma_8$ with the corresponding growth factor. These predictions should be regarded as illustrative, since they rely on simple growth-index parameterizations rather than full modified-gravity calculations.

We also include a projection for the size of the error bars using the DESI BGS DR2 sample, assuming everything about the sample besides the number of objects (which is quadrupled) is unchanged. Provided ours is an accurate estimate of the optical depth, if this slight tension with the $\Lambda$CDM and DGP models persists, we would see it at the $\sim$1.3 and 2$\sigma$ level, respectively, with the BGS DR2 sample. A key improvement for the DR2 analysis would be the accurate and complementary measurement of $\bar \tau$ using a direct observable such as the patchy (anisotropic) screening effect and X-rays.

We emphasize that the results shown here are mostly for proof-of-concept and in the future, we will study the robustness of these results against the method used to estimate $\bar{\tau}$. The main source of uncertainty when estimating it from hydrodynamical simulations stems from two main assumptions: that the galaxy sample selected in the simulation matches the properties of the real data, and that the gas distribution in the simulation at a given redshift and halo mass range reflects accurately the real Universe, including the impact of baryonic feedback. 

Looking ahead, future measurements with DESI DR2 are expected to yield significantly higher SNR (\(\sim 10\sigma\)) for the BGS galaxy sample, as the number of objects will increase by about a factor of four.
This improvement would reduce the fractional uncertainty on \(f\sigma_8^2\) to roughly 10\%, enabling meaningful tests of modified gravity models such as \(f(R)\) and DGP.  
At \(z\sim0.3\), these models predict substantial deviations from GR, so future kSZ measurements will provide a powerful complementary probe to standard clustering and lensing analyses.  

\section{Summary and conclusions}
\label{sec:conc}

In this work, we have presented a new measurement of the pairwise kinematic Sunyaev-Zel’dovich (kSZ) effect using BGS DR1 galaxies overlapping with the ACT DR6 CMB temperature map. This analysis represents one of the first high-significance low-redshift measurements of the kSZ signal, enabled by the combination of large galaxy samples with high-resolution CMB data. Here, we explore the dependence of the signal on galaxy mass and aperture size, validating the robustness of our measurement with a suite of null tests, and then place this result in a cosmological context, demonstrating the viability of using the pairwise kSZ signal as a tracer of cosmic velocity fields at $z \sim 0.3$.

We first examined the dependence of the pairwise kSZ signal on stellar mass, by splitting the BGS sample into low-mass ($M_\star \sim 10^{10} M_\odot$) and high-mass ($M_\star \sim 10^{11.25} M_\odot$) cumulative bins (Fig.~\ref{fig:mass_evolution}). As expected, the signal amplitude increases with stellar mass, consistent with the interpretation that higher-mass galaxies reside in more massive halos with larger gas reservoirs and higher optical depths. The observed scaling is broadly consistent with the hydrodynamical simulation SIMBA, though current uncertainties are too large to allow detailed modeling of feedback effects. This mass dependence underscores the need for accurate stellar-to-halo mass relations and $\tau$ calibrations in order to exploit the cosmological constraining power of the signal.

Next, we studied the impact of varying the aperture photometry scale between $\theta_{\rm ap} = 2.1'$ and $3.5'$ (Fig.~\ref{fig:aperture_evolution}). We find a stable detection across this range, with a peak significance at $\theta_{\rm ap} \simeq 3.3'$, which corresponds closely to the typical closure radius of AGN-active groups \citep{2023MNRAS.524.5391A}. At smaller apertures, the signal is diluted by the filtering of the gas profile, while larger apertures increasingly add noise from uncorrelated structures. The aperture study highlights the sensitivity of the kSZ measurement to halo gas profiles and provides a promising avenue for joint analyses with CMB lensing, which directly probes the halo mass distribution.

To ensure the robustness of our measurement, we performed a series of null tests, including randomizing galaxy positions and temperature decrements (Fig.~\ref{fig:null_tests}). In all cases, the recovered signals were consistent with null, indicating that our detection is not driven by systematics such as map-level artifacts or survey geometry. These null results lend confidence that the observed pairwise kSZ signal is astrophysical in origin, and validate the pipeline for application to deeper datasets in the near future. In Appendix~\ref{app:hilc_vs_single}, we provide evidence for the insensitivity of the estimator to foregrounds such as CIB and tSZ (namely, single-frequency maps recover the same underlying $\bar \tau$ as the hILC map). 

Our measurement is sensitive to the growth of structure through the pairwise velocity field, and in particular constrains the combination $H_0 f\sigma_8^2$ once the optical depth and galaxy bias have been  calibrated. While the current precision and the degeneracy with $\tau$ prevent us from placing tight cosmological constraints, this result demonstrates the feasibility of extracting dynamical information from galaxy–CMB cross-correlations. Combining pairwise kSZ measurements with external probes such as CMB lensing (which constrains halo bias and mass; Fig.~\ref{fig:kappa_profiles}), and hydrodynamical simulations (which provide priors on $\tau$) offers a promising path toward breaking these degeneracies (Fig.~\ref{fig:cosmo_constraints}).

We emphasize that the present work is primarily a proof of concept, establishing the methodology and validating the analysis with BGS DR1 data. With DR2 data, we expect the statistical precision to improve by approximately a factor of two, enabling a decisive step forward in using kSZ velocity measurements to constrain cosmology. In particular, these future analyses will provide new tests of modified gravity models that predict departures from General Relativity in the low-redshift growth rate. Our present results thus represent a crucial milestone toward realizing the full potential of velocity-field cosmology with DESI and upcoming CMB surveys. 

\begin{acknowledgments}

We thank Simone Ferraro and Martin White for enlightening discussions during the preparation of this manuscript.

CS acknowledges support from the Agencia Nacional de Investigaci\'on y Desarrollo (ANID) through Basal project FB210003.
TM acknowledges support from the Agencia Estatal de Investigaci\'on (AEI) and the Ministerio de Ciencia, Innovaci\'on y Universidades (MICIU) Grant ATRAE2024-154740 funded by MICIU/AEI//10.13039/501100011033.

This material is based upon work supported by the U.S. Department of Energy (DOE), Office of Science, Office of High-Energy Physics, under Contract No. DE–AC02–05CH11231, and by the National Energy Research Scientific Computing Center, a DOE Office of Science User Facility under the same contract. Additional support for DESI was provided by the U.S. National Science Foundation (NSF), Division of Astronomical Sciences under Contract No. AST-0950945 to the NSF’s National Optical-Infrared Astronomy Research Laboratory; the Science and Technology Facilities Council of the United Kingdom; the Gordon and Betty Moore Foundation; the Heising-Simons Foundation; the French Alternative Energies and Atomic Energy Commission (CEA); the National Council of Humanities, Science and Technology of Mexico (CONAHCYT); the Ministry of Science, Innovation and Universities of Spain (MICIU/AEI/10.13039/501100011033), and by the DESI Member Institutions: \url{https://www.desi.lbl.gov/collaborating-institutions}. Any opinions, findings, and conclusions or recommendations expressed in this material are those of the author(s) and do not necessarily reflect the views of the U. S. National Science Foundation, the U. S. Department of Energy, or any of the listed funding agencies.

The authors are honored to be permitted to conduct scientific research on I'oligam Du'ag (Kitt Peak), a mountain with particular significance to the Tohono O’odham Nation.

Support for ACT was through the U.S.~National Science Foundation through awards AST-0408698, AST-0965625, and AST-1440226 for the ACT project, as well as awards PHY-0355328, PHY-0855887 and PHY-1214379. Funding was also provided by Princeton University, the University of Pennsylvania, and a Canada Foundation for Innovation (CFI) award to UBC. ACT operated in the Parque Astron\'omico Atacama in northern Chile under the auspices of the Agencia Nacional de Investigaci\'on y Desarrollo (ANID). The development of multichroic detectors and lenses was supported by NASA grants NNX13AE56G and NNX14AB58G. Detector research at NIST was supported by the NIST Innovations in Measurement Science program. Computing for ACT was performed using the Princeton Research Computing resources at Princeton University, the National Energy Research Scientific Computing Center (NERSC), and the Niagara supercomputer at the SciNet HPC Consortium. SciNet is funded by the CFI under the auspices of Compute Canada, the Government of Ontario, the Ontario Research Fund–Research Excellence, and the University of Toronto. We thank the Republic of Chile for hosting ACT in the northern Atacama, and the local indigenous Licanantay communities whom we follow in observing and learning from the night sky.

\end{acknowledgments}

\section*{Data Availability}

All data points shown in the published graph will be also available in a machine-readable form on \url{https://doi.org/10.5281/zenodo.17307201}.

This project uses a script provided in the GitHub repository \url{www.github.com/boryanah/2MPZ_vel/}.

%




\appendix

\section{Comparison of harmonic ILC and single-frequency maps}
\label{app:hilc_vs_single}

An important validation of our analysis is to compare the pairwise signal 
measured from the fiducial sample 
(\(\log (M_\star/M_\odot) > 11\), \(\theta_{\rm ap} = 2.7'\)) 
using the harmonic Internal Linear Combination (hILC) map 
with that obtained from the single-frequency maps directly. 
The hILC combines the ACT DR6 maps centered about 98, 150 and 220 GHz and referred to, respectively, as \texttt{f090}, \texttt{f150}, 
and \texttt{f220}, with the latter downweighted due to strong foregrounds, 
to optimally reduce variance.  
Since the kSZ effect is frequency-independent, one expects the single-frequency 
maps to recover the same signal, albeit with higher variance. 
At the same time, one might worry whether the ILC procedure itself could 
distort the subtle kSZ signal, making this comparison essential.  

\begin{figure*}[t]
    \centering
    \includegraphics[width=0.98\textwidth]{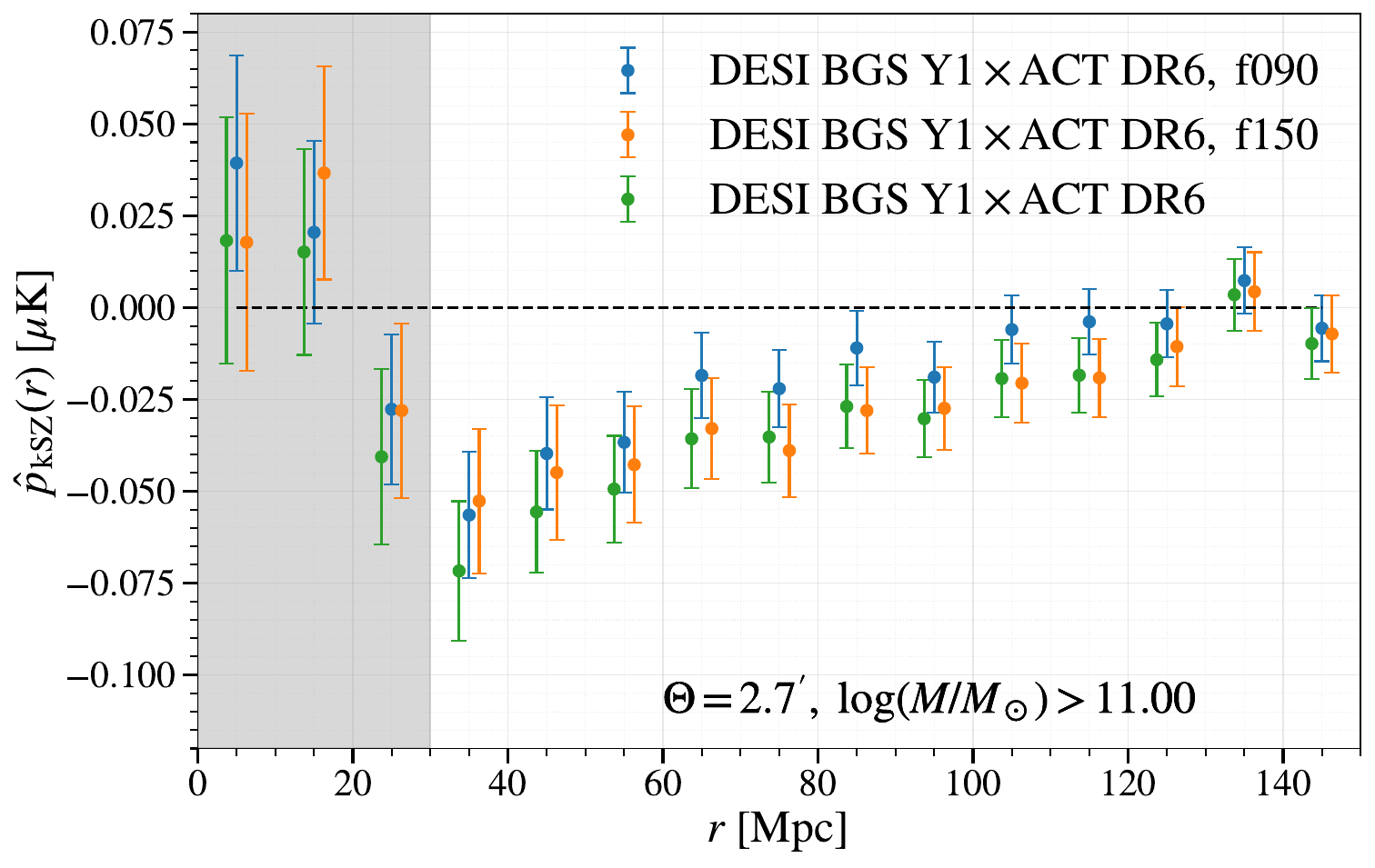}
    \caption{Pairwise kSZ signal for the fiducial sample 
    (\(\log_{10} M_\star > 11\), \(\theta_{\rm ap} = 2.7'\)) 
    measured using the harmonic ILC map and the single-frequency 
    \texttt{f090} and \texttt{f150} maps.  
    All three are consistent within $1\sigma$, with the \texttt{f090} curve 
    slightly lower in amplitude due to its larger effective beam size 
    ($2.1'$) compared to \texttt{f150} ($1.4'$) and hILC ($1.6'$).  
    This behavior is expected and consistent with beam effects seen in 
    stacking analyses of gas profiles.}
    \label{fig:hilc_singlefreq}
\end{figure*}

The results are shown in Fig.~\ref{fig:hilc_singlefreq}. 
We find two key points.  
First, all three measurements (\texttt{hILC}, \texttt{f090}, \texttt{f150}) 
are consistent within $1\sigma$, demonstrating that the hILC procedure 
preserves the kSZ signal.  
Second, the \texttt{f150} and hILC results are more consistent with each other, 
while the \texttt{f090} signal is slightly offset (by $\sim 1\sigma$), 
with a lower optical depth (i.e.\ smaller amplitude, since the signal is 
negative).  
This is naturally explained by the different beam sizes: 
the effective beams are $1.4'$ for \texttt{f150} and $1.6'$ for hILC, 
while \texttt{f090} has a larger beam of $\sim 2.1'$, corresponding to 
stronger smoothing of the CMB map.  
As a result, the optical depth inferred from \texttt{f090} is expected to be 
lower, consistent with trends also seen in gas density profiles from 
stacking measurements (see e.g.\ Refs.~\cite{2025PhRvD.111b3534H}).
We test with the SIMBA simulation that the ratio between $\bar \tau$ estimated from \texttt{f090} and \texttt{f150} from the data is consistent with the simulation finding, validating the robustness of the pairwise kSZ estimator to foreground contamination such as the tSZ and CIB.
On sufficiently large scales (i.e.\ larger than the beam size), these beam 
differences become negligible, and all single-frequency maps yield results in 
perfect agreement.  

In conclusion, this test confirms that the hILC reconstruction performs as 
expected and that the kSZ signal is robust to the choice of temperature map.  

\section{Comparison with SIMBA and Previous BGS Measurements}
\label{app:simba}

In this appendix we compare the stacked gas profiles measured in this work with results obtained in previous analyses as well as with hydrodynamical simulations. Specifically, we consider the cumulative gas profiles (CAP-filtered kSZ measurements) around BGS galaxies with stellar mass thresholds of $\log M_{\ast}/M_{\odot} > 10.5$, obtained in Ref.~\citep{2025PhRvD.111b3534H}, and compare them to a BGS-like galaxy selection defined in the SIMBA simulation. An important difference between the samples is that the present work uses DESI Year~1 BGS data with stellar masses estimated using CIGALE, while Ref.~\citep{2025PhRvD.111b3534H} employed photometric BGS targets with stellar masses from Ref.~\citep{2023AJ....165...58Z}. Nonetheless, given the similarities in target selection, the two samples are expected to be broadly comparable. To construct the BGS-like selection in SIMBA, we utilized the mean halo mass of the photometric BGS sample, $\langle M_h \rangle = 10^{13.03}\,M_{\odot}/h$, as inferred from CMB lensing in Ref.~\citep{2025arXiv250714136H}, to match the halo population.

In particular, since our sample in the data is selected by imposing a stellar mass cut, we mimic this in the simulation by rank ordering the galaxies by their stellar masses. However, the stellar mass estimates from the VACs typically have a scatter of around 0.3--0.5 dex relative to the true stellar mass of the galaxy \citep{2024A&A...691A.308S}, so effectively our sample probes galaxies of lower stellar mass than the imposed minimum. We approximate that by making a cut-off in the stellar mass threshold in the simulations such that the BGS-like sample has a number density of $\bar n_{\rm g} = 5 \times 10^{-3} \ ({\rm Mpc}/h)^{-3}$. Then we impose a maximum halo mass threshold until we match the mean halo mass of the $\log M_{\ast}/M_{\odot} > 10.5$ photometric BGS sample. This halo mass cut aims to approximate the cluster mask applied to our ACT DR6 temperature map when performing the kSZ pairwise measurement. 

\begin{figure}[t]
    \centering
    \includegraphics[width=0.5\textwidth]{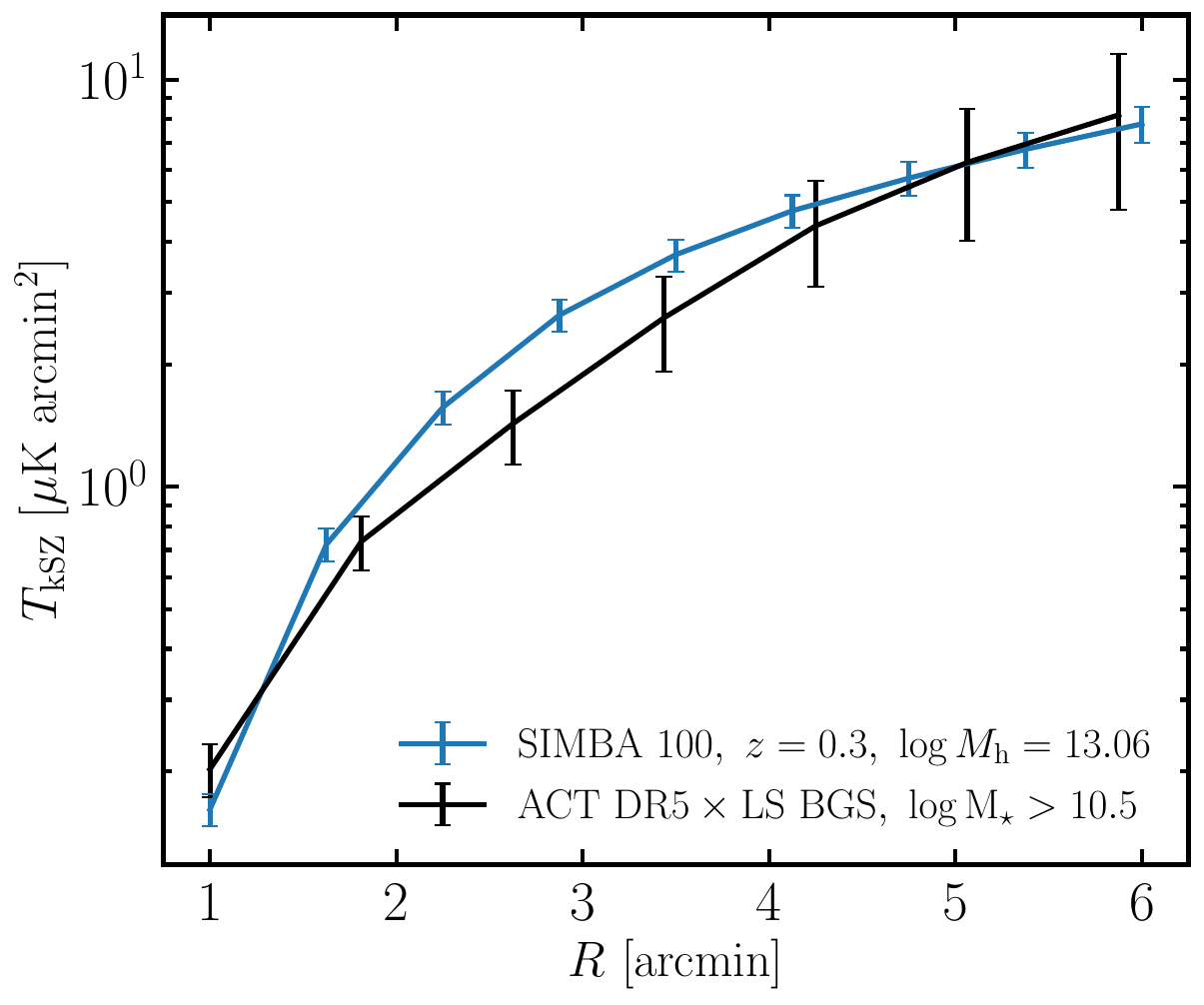}
    \caption{
    Comparison of CAP-filtered kSZ gas profiles around BGS galaxies with $\log M_{\ast}/M_{\odot} > 10.5$. The black points show ACT DR5 measurements using photometric BGS \citep{2025PhRvD.111b3534H}, while the red curve shows the BGS-like sample selected in SIMBA. Both curves are consistent at large scales, as expected from the halo mass matching, while modest deviations appear on small scales, suggestive of differences in feedback efficiency. At $\theta \sim 3.3'$, the two agree within $1\sigma$.}
    \label{fig:simba_comparison}
\end{figure}

Fig.~\ref{fig:simba_comparison} shows the comparison between the ACT DR5 measurements of photometric BGS with $\log M_{\ast}/M_{\odot} > 10.5$ and the BGS-like sample from SIMBA. The curves can be interpreted as the cumulative optical depth profile $\bar{\tau}(\theta)$ as a function of aperture radius. On large angular scales, the agreement is very good, as expected given the halo mass matching. On smaller scales, however, SIMBA deviates somewhat from the measurements, perhaps indicating that its feedback model is not sufficiently strong, despite being stronger than that implemented in TNG300 (which has been shown to underpredict the feedback efficiency of DESI galaxy groups \citep{2024arXiv240707152H,2025arXiv250714136H}). Nevertheless, the differences are at $1\sigma$ around $\theta \approx 3.5'$. Since higher-mass halos are expected to retain a higher fraction of their baryons (and thus to be less sensitive to feedback prescriptions), we anticipate that the agreement between SIMBA and observations will improve further for samples with higher stellar mass thresholds. Although not perfect, this validates our use of SIMBA as a prior on the mean optical depth $\bar{\tau}$ for the $\log M_{\ast}/M_{\odot} > 11$ sample analyzed in the main text. A more detailed comparison with SIMBA and other state-of-the-art simulations is left for future work.

\bibliography{refs,Misha_DESI_supporting_papers2025-05-11}{}
\bibliographystyle{prsty}



\end{document}